\documentclass[conference]{IEEEtran}
\usepackage{microtype}
\pdfoutput=1

\title{Bringing Inter-Thread Cache Benefits to Federated Scheduling --
Extended Results \& Technical Report}

\usepackage{amsthm}
\usepackage{amsmath}
\usepackage{amssymb}
\usepackage{epsfig}
\usepackage{multicol}
\usepackage{xcolor}
\usepackage{graphicx}
\DeclareGraphicsExtensions{.pdf}
\graphicspath{{.},{fig/},{results/},{gnuplot/}}
\usepackage{subcaption}
\usepackage{enumitem}
\usepackage{url}
\usepackage{hyperref}
\usepackage{wrapfig}
\usepackage{framed}
\usepackage[numbers,sort&compress]{natbib}

\usepackage{algorithm,setspace}
\usepackage{algpseudocode}
\usepackage{makecell}

\newtheorem{definition}{Definition}

\newtheorem{obs}{Observation}

\begin{document}

\author{\IEEEauthorblockN{Corey Tessler, Venkata P. Modekurthy, Nathan Fisher and Abusayeed Saifullah} 	Department of Computer Science, Wayne State University, Detroit, MI, USA\\ }

\maketitle

\newlength{\saveisep}
\newlength{\savecsep}
\setlength{\saveisep}{\intextsep}
\setlength{\savecsep}{\columnsep}

\newcommand{\rseps}{
  \setlength{\intextsep}{\saveisep}
  \setlength{\columnsep}{\savecsep}
}
\setlength{\belowcaptionskip}{0pt}
\setlength{\floatsep}{0pt}
\setlength{\textfloatsep}{0pt}
\setlength{\intextsep}{4pt}
\setlength{\dbltextfloatsep}{0pt}
\setlength{\dblfloatsep}{0pt}


\newcommand{\addcite}{({\color{red}\emph{citation}})}

\newcommand{\bundle}{\texttt{BUNDLE}}
\newcommand{\bundlep}{\texttt{BUNDLEP}}
\newcommand{\npmbundle}{\texttt{NPM-BUNDLE}}
%
%
\newtheorem{case}{Case}

\begin{abstract}
  Multiprocessor scheduling of hard real-time tasks modeled by
  directed acyclic graphs (DAGs) exploits the inherent parallelism
  presented by the model. For DAG tasks, a node represents a request
  to execute an object on one of the available processors. In
  one DAG task, there may be multiple execution requests for one
  object, each represented by a distinct node. These distinct
  execution requests offer an opportunity to reduce their combined
  cache overhead through coordinated scheduling of objects as threads
  within a parallel task. The goal of this work is to realize this
  opportunity by incorporating the cache-aware \bundle{}-scheduling
  algorithm into federated scheduling of sporadic DAG task sets.

  This is the first work to incorporate instruction cache
  sharing into federated scheduling. The result is
  a modification of the DAG model named the \emph{DAG with objects and
    threads} (DAG-OT). Under the DAG-OT model, descriptions of nodes
  explicitly include their underlying executable object and number of
  threads. When possible, nodes assigned the same executable object
  are collapsed into a single node; joining their threads when
  \bundle{}-scheduled. Compared to the DAG model, the DAG-OT model
  with cache-aware scheduling reduces the number of cores allocated to
  individual tasks by approximately 20 percent in the synthetic
  evaluation and up to 50 percent on a novel parallel computing
  platform implementation. By reducing the number of allocated cores,
  the DAG-OT model is able to schedule a subset of previously
  infeasible task sets.
  
\end{abstract}

\section{Introduction}

  For hard real-time parallel tasks, where the total execution
  demand of a task may exceed its deadline, federated
  scheduling~\cite{li2014federated, baruah2015federated,
    baruah2015constrained, ueter2018reservation} provides a 
  method for executing each task across multiple cores and an
  accompanying analysis which determines if all tasks will always meet
  their deadlines. To analyze and schedule parallel tasks, each task
  is represented by a directed acyclic graph (DAG).

  Nodes within a DAG represent the release and complete execution
  of an object upon one of the ${m}$ identical cores of the
  system. Edges between nodes indicate precedence constraints between
  nodes; a node may not begin executing until all predecessors have
  completed. Associated with every node is a worst-case execution
  time (WCET) bounding any complete execution. Schedulability analysis
  of a task's DAG depends on the task's workload (sum of WCETs of all
  nodes) and critical path length (path through the DAG with greatest
  WCET).

  Worst-case execution time calculation accounts for architecture
  features including cache memory. The variability in execution times
  due to cache memory has been well studied for uniprocessor
  single-threaded task sets in works such
  as~\cite{Tomiyama:2000, tan:2007, Lee:1998,
    altmeyer:2011, negi:2003, altmeyer:2012, Li:2009, zhang2016cache,
    chattopadhyay2014cache}. Scheduling of tasks on multi-processor
  systems with cache memory has been studied in works such
  as~\cite{Liu:2016,Saeeung:2010, Cole:2011, Xu:2016, Xiao:2017,
    Calandrino:2009b, Calandrino:2008}. 
  In most previous work on both multiprocessor and uniprocessor
  real-time systems, cache memory contributes primarily
  negatively to schedulability by increasing WCET values. Preemptions
  between jobs introduce cache-related preemption delays (CRPD) for both
  uniprocessor and multi-processor systems. Multi-processor
  multi-threaded systems
  with shared caches are also affected by evictions from concurrent
  execution as well as cache coherency delays across cores
  ~\cite{Calandrino:2008, Calandrino:2009b, Xiao:2017, Xu:2016,
    Saeeung:2010, Cole:2011}. The method proposed in this work is the
  first to incorporate instruction cache reuse beneficially into real-time
  scheduling decisions for federated scheduling of multi-processor
  multi-threaded systems.

  In the setting of uniprocessor multi-threaded task systems, the
  \bundle{}-based
  approaches~\cite{tessler:2016,tessler:2018,tessler:2019} (referred to
  as \bundle{} throughout the rest of this work) treat cache memory
  positively, creating a benefit to schedulability. \bundle{}
  restricts the execution of the multiple threads of task on a single
  processor in a cache cognizant manner. This restricted execution
  allows the sharing of cached values to be quantified as the
  inter-thread cache benefit. The accompanying \bundle{} analysis
  incorporates the inter-thread cache benefit into a WCET function for
  each task. These WCET functions accept the number of threads
  released with each job and quantifies the total benefit of
  ``bundling'' the threads together.

  \bundle{} is limited to single processor multi-threaded tasks. The
  inter-thread cache benefit applies exclusively to instruction
  caches. Furthermore, \bundle{}'s scheduling and WCET analysis
  techniques are limited to a single executable object. As such,
  \bundle{} is not directly applicable to parallel DAG tasks utilized
  by federated or global multi-processor schedulers.

  This work incorporates cache memory positively into multi-processor
  parallel tasks by joining \bundle{}'s analysis and scheduling
  techniques to those of federated scheduling. This is achieved by
  treating executable objects (nodes) of parallel DAG tasks as threads
  scheduled by \bundle{}. Each individual node of a DAG task
  represents a single thread of execution of the underlying
  object. Nodes sharing the same underlying object may be
  \emph{collapsed} into a single node and the combined threads
  scheduled by \bundle{}.

  The purpose of collapse is to increase schedulability by reducing
  the number of processors dedicated to high utilization
  tasks. However, collapse may not be arbitrarily applied to nodes of
  the same object. There are several challenges when considering
  collapsing two nodes, doing so may:
  \begin{itemize}
  \item Introduce a cycle to the DAG
  \item Produce an infeasible parallel task
  \item Increase the number of cores allocated to a task
  \end{itemize}

  To achieve the goal of reducing
  the number of cores allocated to high utilization tasks while
  carefully selecting which nodes to collapse the following
  contributions are made:

\begin{itemize}
\item A modification to the DAG model named DAG-OT, the first parallel
  task model to include inter-thread cache benefits
\item The concepts of collapse and collapse candidacy
\item An algorithm for collapsing nodes
\item Heuristics for ordering collapse of nodes within a task
\item An evaluation of synthetic tasks demonstrating the benefits of
  collapse and \bundlep{} scheduling of nodes under a federated
  scheduler
\item A feasibility study with a parallel DAG-OT task scheduler
  operating on physical hardware demonstrating the potential cache
  benefits 
\end{itemize}

These contributions are made in the following
sections. Section~\ref{sec:related} expands upon the related
work. Section~\ref{sec:model} describes \bundle{}-based scheduling,
the existing DAG model, and the proposed DAG-OT model.
Section~\ref{sec:collapse} describes the collapse operation and its impact.
Section~\ref{sec:dagot-sched} introduces the general algorithm for
collapsing nodes. Section~\ref{sec:selection} describes the proposed
heuristics used to order candidates for collapse. 
Section~\ref{sec:lowutil-collapse} describes the collapse of
low utilization tasks. Section~\ref{sec:sched-analysis} describes
the schedulability for tasks of the proposed
model. Section~\ref{sec:evaluation} describes the methods, metrics,
and results of the synthetic evaluation. Section~\ref{sec:poc}
presents the feasibility study and
results. Section~\ref{sec:conclusion} concludes the work.

\section{Related Work}
\label{sec:related}

Parallel hard real-time DAG tasks may be scheduled by
federated~\cite{li2014federated, baruah2015federated,
  ueter2018reservation, melani2015response, sun2018capacity}
or global~\cite{saifullah2013multi, saifullah2014parallel,
  lakshmanan2010scheduling, baruah2015global} 
policies. Federated scheduling improves the analytical bounds of
global scheduling by dedicating cores to 
tasks that require more than one core to meet their
deadlines. 

For multi-processor systems, the impact of cache memory focuses
  on shared caches. When caches are shared between cores, an object
  executing on one core may evict values placed there by another
  object on a distinct core -- increasing execution times. There are
  several works dedicated to mitigating or providing bounds on the
  number of evictions with global scheduling policies,
  including~\cite{Calandrino:2008, Calandrino:2009b, Xiao:2017,
    Xu:2016}. It should be noted the tasks in~\cite{Xiao:2017,
    Xu:2016} are not parallel tasks. Cache coherency and false
  sharing~\cite{Cole:2011, Saeeung:2010, Liu:2016} is an another
  source of execution time extension for parallel tasks running on a  
  multi-processor system with shared caches.

In the setting of uniprocessor systems executing
  single-threaded tasks, cache memory has been well studied. WCET
analysis accounting for cache reuse of single-threaded tasks and
direct map caches is presented in~\cite{Arnold:1994, Mueller:1995,
  Mueller:2000} and expanded to set-associative caches
in~\cite{Li:1996}. Addressing the impact of cache memory in a
preemptive setting is the purpose of CRPD
analysis~\cite{Tomiyama:2000, tan:2007, Lee:1998, staschulat:2005,
  altmeyer:2012, ju:2007, lunniss:2014a, lunniss:2013, lunniss:2016,
  lunniss:2014b}. 

Each of the CRPD analytical methods seek to accurately estimate the
impact of preemptions upon the WCET bound for single-threaded
tasks. Other approaches seek to mitigate CRPD's run-time impact. The
PREM~\cite{Bak:2012, Pellizzoni:2011, Ward:2013}
model divides tasks into load and execute phases, preventing cross
tasks cache interference. Explicit preemption point
placement~\cite{Simonson:1995, Wang:1999, Lee:1998, Bril:2017,
  Bertogna:2011} limits when a job may preempt another based upon the
cache impact it may have.

Caches receive a common treatment in the uniprocessor works related to
CRPD and multi-processor works related to shared caches. Specifically,
caches are seen from the negative perspective, exclusively detracting
from schedulability analysis. The only works the authors' are aware of
that take a positive perspective are \emph{persistent cache
  blocks}~\cite{Rashid:2016,Rashid:2017} in the uniprocessor
single-threaded setting and \emph{cache spread}~\cite{Calandrino:2009}
in the multi-processor setting utilizing global scheduling.

  The work proposed herein applies to the federated scheduling of hard
  real-time parallel DAG tasks on multi-processor systems, focusing
  on the impact of scheduling decisions in the presence of dedicated
  (not shared) caches. To the authors knowledge, there are no existing
  works that address this setting. Furthermore, the proposed approach
  treats instruction caches positively by decreasing the number of
  cores dedicated to high utilization tasks in an attempt to increase
  schedulability.

A positive perspective of caches is taken
 by \bundle{}~\cite{tessler:2016,tessler:2018,tessler:2019} for
 multi-threaded tasks running on a single processor. This positive
 perspective is reflected in the WCET of a task which includes
 the \emph{inter-thread cache benefit}: the speed-up one thread
 experiences by another placing values in the cache. The benefit is
 restricted to instruction caches when threads are scheduled by
 the \bundle{} scheduling algorithm.

\bundle{} analysis and scheduling are central to the combined
scheduling approach proposed in Section~\ref{model:dag-ot}, depending
upon the \bundle{} calculated WCET values for collapse decisions.
As a product of \bundle{} analysis, each multi-threaded task is
assigned a worst-case execution time and cache overhead (WCETO)
function ${c(\eta):\mathbb{N}^+ \rightarrow \mathbb{R}}$. Where ${c(\eta)}$
is an upper bound on the amount of time required to execute ${\eta}$
threads by \bundle{}, which encapsulates the inter-thread cache
benefit. The result is a strictly increasing concave function 
with respect to ${\eta}$. Summarily, for ${\eta+1}$ threads,
${c(\eta+1) \le c(\eta) + c(1)}$. By combining \bundle{} and
federated scheduling the concave property of \bundle{} analysis can be
leveraged to increase the schedulability of parallel tasks.

\section{System Model}
\label{sec:model}
This work proposes changes to the parallel DAG
  model~\cite{li2014analysis} of hard real-time tasks to support
  collapse operations. The decision to collapse nodes is
  a scheduling decision that depends upon the \bundle{}'d execution of
  combined threads upon a single core. The purpose of this section is
  to describe the \bundle{} model in Subsection~\ref{model:bundle},
  summarize the DAG model in Subsection~\ref{model:dag} and federated
  scheduling in Subsection~\ref{model:federated}, describe the
  proposed model which combines federated and \bundle{} scheduling in
  Subsection~\ref{model:dag-ot}, and lastly illustrate the impact of
  collapse under the combined model.
\subsection{\bundle{}}\label{model:bundle}
Tasks in the 
\bundle{}~\cite{tessler:2016,tessler:2018,tessler:2019} model are
represented by a tuple ${\tau_i = (p_i, d_i, c_i(\eta_i),
  \eta_i, o_i)}$. A task has the familiar minimum inter-arrival time ${p_i}$
and relative deadline ${d_i}$. A task has an underlying executable
object ${o_i}$, and a number of threads
released per job ${\eta_i}$. With each job release of ${\tau_i}$,
${\eta_i}$ threads are simultaneously released and must complete
before the relative deadline ${d_i}$. The task's WCETO is given by
${c_i(\eta_i)}$ which provides an upper bound to complete all
${\eta_i}$ threads when scheduled in \bundle{}'s cache cognizant
manner.

To produce the WCETO for a task, the control flow graph of the
executable object is divided into conflict free regions: sub-graphs of
the control flow graph where no two instructions map to the same cache
block. Conflict free regions serve as input to the \bundle{}
scheduling algorithm, which maximizes the number of threads executing
over each region~\cite{tessler:2018} in turn maximizing the
inter-thread cache benefit.

Analysis of jobs scheduled by \bundle{} incorporates the inter-thread
cache benefit in the task's WCETO function
${c(\eta)}$ for ${\eta \in \mathbb{N}^+}$ threads. Every increase in
${\eta}$ maximizes the contribution of an individual thread. The
result is that ${c(\eta)}$ is a discrete concave function. Consider
the addition of a single thread ${c(\eta + 1)}$ compared to the
addition of two threads ${c(\eta + 2)}$. The WCETO increase of
${c(\eta)}$ to ${c(\eta + 1)}$ must be greater than or equal to the
increase from ${c(\eta + 1)}$ to ${c(\eta + 2)}$: ${c(\eta + 1) -
  c(\eta) \ge c(\eta + 2) - c(\eta + 1)}$. If it were not, the
increase of ${c(\eta + 1)}$ would not be maximal. Furthermore, if the
increase of one thread were less than that of a second thread the
bound of ${c(\eta + 1)}$ would be optimistic and unsafe. Thus, for any
${\eta_a < \eta_b < \eta_c}$ the point ${(\eta_b, c(\eta_b))}$ lies
above the line defined by ${(\eta_a, c(\eta_a))}$ and ${(\eta_c,
  c(\eta_c))}$, therefore ${c(\eta)}$ is concave. Multi-threaded
programs executed by \bundlep{}~\cite{tessler:2018} illustrate the
discrete concave growth described by the analysis.

\subsection{DAG Model}\label{model:dag}
Tasks in the parallel DAG model~\cite{li2014analysis} are
  represented by a tuple ${\tau_i = (T_i, D_i, G_i)}$ of minimum inter
  arrival time ${T_i}$, implicit deadline ${D_i = T_i}$ and directed acyclic
  graph ${G_i = (V_i, E_i)}$. The set of ${n}$ tasks is given by
  ${\tau = \{\tau_1, \tau_2, ..., \tau_n\}}$. The set of all DAGs is
  denoted ${\mathbb{G} = \{G_1, G_2, ..., G_n\}}$.

Within a DAG ${G_i}$, a node ${v \in V_i}$ represents the
execution of a single thread. A 
thread executes on exactly one of the ${m}$ cores of the target
architecture (or distributed system). Each node is implicitly
associated with an underlying executable object ${\alpha_v}$: a set of
machine instructions reachable from a single entry point. A worst-case
execution ${c_v}$ time is associated with every node ${v}$; an upper
bound on the execution time required to complete the thread without
interruption on a single core. An edge ${(u,v) \in E_i}$ indicates an
execution dependency between ${u,v \in V_i}$. For ${v}$ to begin
execution on any core, all immediate predecessors ${\{u|(u,v) \in
  E_i\}}$ must run to completion.   

For simplicity of analysis, every DAG ${G_i}$ must have exactly one
source and sink node, ${s,t \in V_i}$ respectively. A source ${s}$ has
no incoming edges, ${\not \exists u~|~(u,s) \in E_i}$. A sink ${t}$
has no outgoing edges, ${\not \exists v~|~(t,v) \in E_i}$. Without
loss of generality, when a DAG contains multiple sources, the DAG is
augmented by adding an ``empty source'': a single node with zero
execution cost that is connected by outgoing edges to existing
sources. Similarly, for a DAG with multiple sinks an ``empty sink'' is
added with zero execution cost connected by incoming edges from the
existing sinks. 

\begin{wrapfigure}[8]{r}[0pt]{0.4\columnwidth}
  \centering
  \includegraphics[width=0.25\columnwidth]{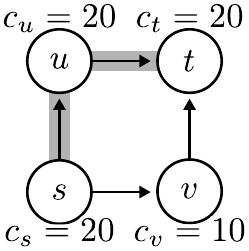}
  \caption{A DAG Task}
  \label{fig:dag-ex}
\end{wrapfigure}

Jobs of a task begin with one thread of ${s}$ on one core. Jobs
terminates when the single thread of ${t}$ completes. During the
execution of a job, up to ${m}$ cores may execute any of the ${v \in
  V}$ threads in parallel. A task ${\tau_i \in \tau}$ generates a
potentially infinite number of jobs, each arriving no less than
${T_i}$ time units apart. All jobs of ${\tau_i}$ must complete within
${D_i = T_i}$ time units.

An example DAG task is shown in Figure~\ref{fig:dag-ex}. Accompanying
each node is a single-threaded WCET. For ${u}$ and ${v}$, their WCET
values are ${c_u = 20}$ and ${c_v = 10}$ respectively. Edges
illustrate the dependency order of execution, such as ${(s,v)}$
precluding ${v}$ from executing until ${s}$ has completed. 

For a DAG ${G_i = (V_i, E_i)}$, the length of a path through the graph
is the sum of WCET values of all nodes along the path. The
\emph{critical path} ${\lambda_i}$ of ${G_i}$, is a path from ${s}$ to
${t}$ with the greatest length ${L_i}$ called the
\emph{critical path length}. If there are multiple paths with equal length
${L_i}$, only one is selected as the critical path.  The
\emph{workload} of ${G_i}$ is the sum of all WCET values
${v \in V_i}$. \emph{Utilization} of the task ${\tau_i}$ is the ratio
of its workload and minimum inter-arrival time.

\begin{table}[ht]
  \centering
  \begin{tabular}{|ll|}
    \hline
    \makecell{Critical Path Length of ${G_i}$ \\
    \parbox{.4\columnwidth}{
        \begin{equation}
          \label{eq:criticalpath}
          L_i = \sum_{v \in \lambda_i} c_v
        \end{equation}
    }}
    &
    \makecell{Workload of ${G_i}$ \\
    \parbox{.4\columnwidth}{
        \begin{equation}
          \label{eq:workload}
          C_i = \sum_{v \in V_i} c_v
        \end{equation}
    }} \\
    \makecell{Utilization of ${G_i}$ \\
    \parbox{.4\columnwidth}{
        \begin{equation}
          u_i = C_i / T_i
        \end{equation}
    }}
    &
    \makecell{Utilization of ${\tau}$ \\
    \parbox{.4\columnwidth}{
        \begin{equation}
          U = \sum_{\tau_i \in \tau} u_i
        \end{equation}
    }}
    \\
    \hline
  \end{tabular}
  \caption{Definitions for Parallel DAG Task Sets}
\end{table}

In Figure~\ref{fig:dag-ex}, the critical path 
${\lambda = \langle s, u, t \rangle}$ is highlighted. The calculated
critical path length is ${L = c_s + c_u + c_t = 60}$ and workload
${C = c_s + c_u + c_v + c_t = 70}$.

\subsection{Federated Scheduling}
\label{model:federated}
\setlength{\FrameSep}{1pt}
\begin{wrapfigure}[6]{r}[0pt]{0.4\columnwidth}
  \begin{framed}
  \begin{equation}\label{eq:m}
      m_{i} = \left\lceil \frac{C_{i} - L_{i}}{D_{i} - L_{i}}
      \right\rceil
  \end{equation}
  \vspace*{-0.5\baselineskip}\hspace{1pt}
  \caption{${m_i}$ for ${\tau_i \in \tau_{high}}$}%
  \end{framed}
\end{wrapfigure}

\emph{Federating scheduling}~\cite{li2014analysis} is a partitioned
scheduling algorithm variant and analysis method developed for
parallel DAG task sets. It divides the task set ${\tau}$ into two
disjoint sets. Tasks with utilization greater than one are placed in the
\emph{high utilization task set} ${\tau_{high}}$. The
\emph{low utilization task set} ${\tau_{low}}$ contains the remainder
of ${\tau}$. Every task ${\tau_i}$ of ${\tau_{high}}$ is assigned
${m_i}$ dedicated cores, where ${m_i}$ is given by 
Equation~\ref{eq:m}. Only threads of ${\tau_i}$ may execute on the
${m_i}$ cores dedicated to it. All jobs of a high utilization task 
${\tau_i}$ scheduled on ${m_i}$ cores are guaranteed to meet their
deadlines~\cite{li2014analysis}.

The number of cores allocated to all high utilization tasks is denoted 
 $m_{high} = \sum_{\tau_{i} \in \tau_{high}} m_{i}$. 
The remaining cores of low utilization tasks are denoted
${m_{low} =  m - m_{high}}$. A task set ${\tau}$ is schedulable under
federated scheduling if ${m_{low}}$ is non-negative and all tasks of
${\tau_{low}}$ are partitioned on the ${m_{low}}$ processors while
meeting their deadlines when threads within jobs are scheduled
sequentially.  

Any greedy, work-conserving, parallel scheduler may be used to
schedule a high utilization task ${\tau_i \in \tau_{high}}$ on its
${m_i}$ dedicated cores. Low utilization tasks are treated as
sequential tasks, executing at most one thread of a job at a time. Any
multiprocessor scheduling algorithm (such as partitioned EDF) may be
used to schedule all the low utilization tasks on the $m_{low}$
allocated cores.

\subsection{Proposed Model: DAG-OT}\label{model:dag-ot}
The model proposed in this work augments the existing DAG model
  by explicitly including the implicit number of threads and
  executable objects associated with every node ${v \in
    V}$. Doing so requires modification to the WCET of a node,
  converting the static value ${c_v}$ to a function in terms of the
  number of threads executed. For clarity, the existing
  model is referred to as the directed acyclic graph model of parallel
  tasks or simply ``the DAG model'', the proposed model is named the
  DAG with objects and threads or ``the DAG-OT model''. 

For a DAG in the DAG model ${G_i = (V_i, E_i)}$, two distinct nodes
  ${u, v \in V_i}$ represent the release of one thread of execution
  over their underlying executable objects ${\alpha_u}$ and
  ${\alpha_v}$. There is no restriction on the relationship between
  ${\alpha_u}$ and ${\alpha_v}$, they may be distinct or identical
  objects. The first proposed change to the DAG model is to explicitly
  include the executable object in a node's description.

Similarly, for a node ${v \in V_i}$ in the DAG model, the
  execution of a single thread is bounded by a single WCET value
  ${c_v}$. The second proposed change to the DAG model is to
  explicitly include the number of threads ${\eta_v}$ and present the
  WCET of a node as function in terms of the number of
  threads executed ${c_v(\eta):\mathbb{N}^+ \rightarrow
    \mathbb{R}^+}$.

\begin{figure}[ht]
  \centering
  \begin{subfigure}[b]{0.4\columnwidth}{
      \includegraphics[width=\textwidth]{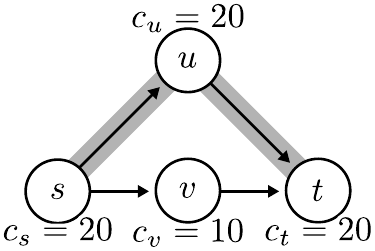}
      \caption{DAG model}
      \label{fig:dag-change-01}
    }
  \end{subfigure} \quad
  \begin{subfigure}[b]{0.4\columnwidth}{
      \includegraphics[width=\textwidth]{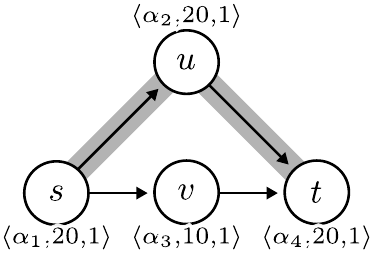}
      \caption{DAG-OT Model}
      \label{fig:dag-change-02}
    }
  \end{subfigure}
  \caption{From DAG to DAG-OT}
  \label{fig:dag-change}
\end{figure}

Combining the proposed changes, a node ${v \in V_i}$ in the DAG-OT
model is represented by a tuple ${v = \langle \alpha_v, c_v(\eta), \eta_v
  \rangle}$. Figure~\ref{fig:dag-change} presents the differences between  
the DAG and DAG-OT models visually. A consistent illustrative
shorthand is used in this work, with the order of nodes tuple's
preserved and the critical path highlighted in gray.

Nodes of the DAG-OT model are compatible with nodes of the DAG
model \cite{li2014analysis}, where nodes from the DAG model can be 
expressed as ${v = \langle \alpha_{v}, c_{v}(\eta), \eta_v=1
  \rangle}$ under DAG-OT. This is illustrated by
Figures~\ref{fig:dag-change-01} and ~\ref{fig:dag-change-02}, which
are equivalent. 

The motivation for including the executable object, threads, and
  WCET as a function in the description of a node is to satisfy the
\bundle{} model and facilitate the combined scheduling
technique. Combining the federated and \bundle{} scheduling
techniques, each node is treated as single unit of execution to be
\bundle{} scheduled upon one core. Each node requires ${\eta_v \ge 1}$
threads of object ${\alpha_v}$ to be executed by \bundle{}.

Under the DAG-OT model, when a node ${v \in V_i}$ is selected
  for execution all ${\eta_v}$ threads of the object ${\alpha_v}$ are
  executed and scheduled by \bundle{} on one core. The total execution
  required to complete all threads is bounded by the WCETO function
  provided by \bundle{} analysis and associated with the node as
  ${c_v(\eta)}$.

Under federated scheduling (and in this work) DAG tasks execute on a
parallel system with ${m}$ identical cores. Requiring uniform
cores ensures the validity of the WCET bound for each node regardless
of which core the thread executes upon. Furthermore, this work requires each
core to have identical cache configurations (hierarchy, size,
etc.), memory architecture, and be timing-compositional~\cite{Hahn:2015}.
Doing so guarantees the worst-case execution time and cache
overhead (WCETO) of every node will be consistent across all
cores. \bundle{} WCETO analysis is limited to the per-core dedicated
instruction caches. Data caches, and cache memory shared between cores
are not considered and are reserved for future work.

For the DAG-OT model, the definitions of critical path length and
workload must be updated, given by
Equations~\ref{eq:dag-ot-critical-path-length} and
\ref{eq:dag-ot-workload}.

\begin{definition}[DAG-OT Critical Path Length of ${G_i}$]
  \begin{equation}
    \label{eq:dag-ot-critical-path-length}
    L_i = \sum_{v \in \lambda_i} c_v(\eta_v)
  \end{equation}
\end{definition}

\begin{definition}[DAG-OT Workload of ${G_i}$]
  \begin{equation}
    \label{eq:dag-ot-workload}
    C_i = \sum_{v \in V_i} c_v(\eta_v)
  \end{equation}
\end{definition}

\subsection{Growth Factors}

For simplicity of presentation and analysis the WCETO function
${c_u(\eta)}$ for a node ${u}$ is described by a \emph{growth factor}
${\mathbb{F}_u}$. A growth factor upper bounds the discrete concave
WCETO of a node by a linear function using the single threaded
(${c_u(1)}$) value at the cost of increased pessimism. This
simplification is leveraged in the evaluation. 

\begin{definition}[Growth Factor ${\mathbb{F}}$]
  For a node ${u \in V}$ of a DAG ${G_i = (V, E)}$, the \emph{growth factor}
  of ${u}$ is a value ${\mathbb{F}_u \in (0,1]}$ that satisfies
  Equation~\ref{eq:factor} for all ${\eta_u \ge 1}$.
  \begin{equation}
    \label{eq:factor}
    c_u(\eta_u) \le c(1) + \mathbb{F}_u \cdot (\eta_u - 1) \cdot c_u(1)
  \end{equation}
\end{definition}

\begin{figure}[ht]
\centering
  \includegraphics[width=0.8\columnwidth]{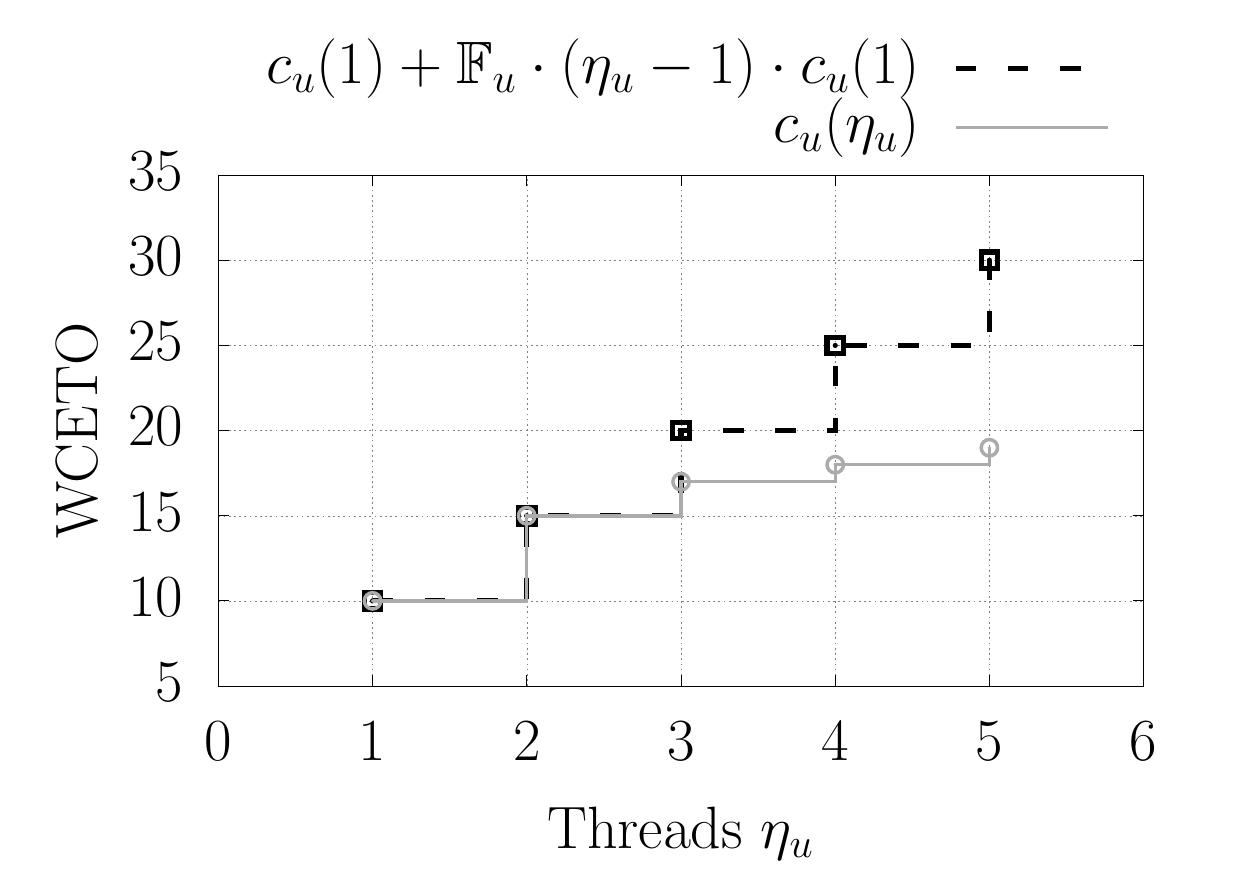}
  \caption{Example Growth Factor}
  \label{fig:growth-factor}
\end{figure}

An example for a node ${u}$, associated ${c_u(\eta_u)}$, and growth
factor ${\mathbb{F}_u = .5}$ is shown in
Figure~\ref{fig:growth-factor}. The values of ${c_u(\eta_u)}$ are
${10, 15, 17, 18, 19}$ for ${\eta_u \in [1, 5]}$. While any growth
factor greater than .5 would satisfy the definition, the minimum was
selected for illustrative purposes.

\section{Collapsing Nodes}
\label{sec:collapse}

To bring the inter-thread cache benefit to the DAG-OT model, this work
proposes the concept of \emph{collapsing} nodes. Under the DAG-OT
model, two nodes ${u,v \in V_i}$ which execute the same object
${\alpha_u = \alpha_v}$ may potentially be combined into a single
node are referred to as \emph{candidates}. Candidates feature prominently
in the fork-join~\cite{lui:1998} and MapReduce~\cite{MapReduce}
parallel task models; which are restricted instances of parallel DAG
tasks. A general DAG task may include fork-join or MapReduce
sub-graphs as part of the task's graph. Collapsing two nodes into a
single node turns two distinct execution requests executing on
(possibly) distinct cores, into a single request to execute the
combined threads on one core using \bundle{} scheduling. By virtue of
\bundle{}'s analysis incorporating the inter-thread cache benefit, the
WCETO of the combined node may be less than the sums of the individual
nodes.

\begin{definition}[Candidate for Collapse]
  \label{def:candidates}
  For a DAG ${G_i = (V, E)}$ and nodes ${u,v \in V}$, ${u}$ and ${v}$
  are \emph{candidates} for collapse if and only if they share an
  executable object ${\alpha_u = \alpha_v}$. 
\end{definition}

\begin{figure}[ht]
  \captionsetup[subfigure]{justification=centering}
  \centering
  \begin{subfigure}[b]{0.4\columnwidth}{
      \centering
      \includegraphics[height=.6\textwidth]{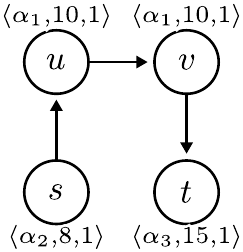}
      \caption{Pre-Collapse}
      \label{fig:serial-ex-01}
    }
  \end{subfigure}
  \begin{subfigure}[b]{0.4\columnwidth}{
      \centering
      \includegraphics[height=.6\textwidth]{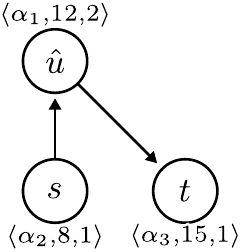}
      \caption{Post-Collapse}
      \label{fig:serial-ex-02}
    }
  \end{subfigure}
  \caption{Node Collapse}
  \label{fig:col-serial-01}
\end{figure}

To illustrate, consider Figure~\ref{fig:col-serial-01}. Nodes ${u}$
and ${v}$ share the same executable object ${\alpha_1}$. If the WCETO of
one thread scheduled by \bundle{} on one core is 10 and two is
12, two nodes executing on separate cores will require 20 total
cycles. Collapsing ${u}$ and ${v}$, requiring the two threads to be
scheduled in a cache cognizant manner on one core by \bundle{} reduces
the workload (and potentially the critical path length) by 8.
  
Collapse restricts the execution of threads and cores. In
Figure~\ref{fig:col-serial-01} pre-collapse ${u}$ and ${v}$ may have
executed on distinct cores. Post-collapse the combined threads of
${u}$ and ${v}$ must execute on the same core scheduled by \bundle{}. To
differentiate between pre and post-collapse values a ``hat'' will be
used for the latter. In Figure~\ref{fig:col-serial-01}, before
collapse ${u}$ and ${v}$ each execute one thread. Collapsing the two
nodes into ${\hat{u}}$ joins the two threads ${\eta_{\hat{u}} = 2 =
  \eta_u + \eta_v}$. The pre-collapse workload is ${C_i = 43}$ and
post-collapse workload ${\hat{C}_i = 35}$. The reduction in workload
is due to the concave WCETO function
${c_u(\eta) = c_v(\eta) = c_{\hat{u}}(\eta)}$, where ${c_u(1) = 10}$ 
and ${c_u(2) = 12}$.

Formally, the collapse operation is defined as follows. 
\begin{definition}[Collapse ${\hat{u} \gets u \Join v}$] For
  pre-collapse nodes ${u, v \in V}$ of ${G_i = (V,E)}$, collapsing ${u}$ and
  ${v}$ (denoted ${u \Join v}$) into ${\hat{u}}$, resulting in a new DAG named
  ${\hat{G}_i = (\hat{V}, \hat{E})}$ where:
  {\small
  \begin{align}
    \eta_{\hat{u}} &\gets \eta_u + \eta_v \label{eq:col_t} \\
    \alpha_{\hat{u}} &\gets \alpha_u \label{eq:col_a} \\
    c_{\hat{u}} &\gets c_u \label{eq:col_c} \\
    \hat{V} &\gets \hat{u} \cup V \setminus \{u, v\} \label{eq:col_v} \\
    \hat{E} &\gets \label{eq:col_e}
    \{(\hat{u}, y) | (u, y) \in E \lor (v, y) \in E)\} \\
    &\cup \{(x, \hat{u}) | (x, u) \in E \lor (x, v) \in E)\} \nonumber
    \\
    &\cup E \setminus \{(x,y) | x \in \{u,v\} \lor y \in \{u,
    v\}\} \nonumber 
  \end{align}
  }
\end{definition}

Equation~\ref{eq:col_t} joins the threads of ${u}$ and ${v}$ to
${\hat{u}}$. Equation~\ref{eq:col_a} and~\ref{eq:col_c} assigns the
executable object and WCETO function shared between ${u}$ and ${v}$ to
${\hat{u}}$. Equation~\ref{eq:col_v} takes the nodes from ${G_i}$ and
copies them to ${\hat{G}_i}$, removing ${u}$ and ${v}$. Similarly,
Equation~\ref{eq:col_e} copies the edges of ${G_i}$ to ${\hat{G}_i}$
while removing incoming and outgoing edges of ${u}$ and ${v}$
replacing them with incoming and outgoing edges of ${\hat{u}}$.

\subsection{Infeasibility and the Impact of Collapse}
\label{sec:critical-path-reduction}

\begin{figure}[ht]
  \centering
  \begin{subfigure}[b]{0.4\columnwidth}{
      \centering
      \includegraphics[height=.6\textwidth]{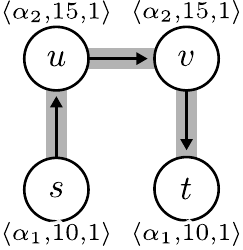}
      \caption{Pre-Collapse}
      \label{fig:collapse-reduce-01}
    }
  \end{subfigure}
  \begin{subfigure}[b]{0.4\columnwidth}{
      \centering
      \includegraphics[height=.6\textwidth]{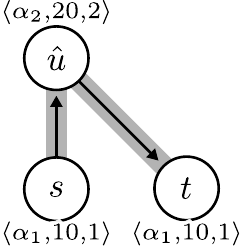}
      \caption{Post-Collapse}
      \label{fig:collapse-reduce-02}
    }
  \end{subfigure}
  \caption{Critical Path Reduction}
  \label{fig:collapse-reduce}
\end{figure}

Collapsing nodes may reduce the critical path length ${L_i}$. This is
illustrated by Figure~\ref{fig:collapse-reduce}, where the
pre-collapse critical path length is ${L_i = 50}$. After collapsing
${\hat{u} \leftarrow u \Join v}$, the critical path length of
${\hat{G}_i}$ is ${\hat{L}_i = 40}$.

\begin{obs}[Critical Path Reduction]
  \label{obs:critical-path-reduction}
  For a DAG ${G_i = (V, E)}$ and candidate nodes ${u, v \in V}$, the
  collapse of ${u \Join v}$ into ${\hat{u}}$ may reduce the critical
  path length in ${\hat{G}_i}$: ${\hat{L}_i \le L_i}$.
\end{obs}

Under the DAG model, a task ${\tau_i}$ is infeasible (for any number of
dedicated cores ${m_i}$) if the critical path length is greater than
the deadline, i.e., ${L_i > D_i}$. A task ${\tau_i}$ deemed infeasible
due to critical path length and period under the DAG model (${L_i > D_i}$)
may become feasible (and possibly schedulable) under the DAG-OT model
by collapse and Observation~\ref{obs:critical-path-reduction}
(${\hat{L}_i \le D_i}$). Thus the ${L_i > D_i}$ infeasibility test
does not apply pre-collapse to the DAG-OT model. However, for any
post-collapse ${\hat{G}_i}$ of ${\tau_i}$ if ${\hat{L}_i > D_i}$ the
task set is unschedulable under DAG-OT.

\begin{obs}[Critical Path Extension]
  \label{obs:critical-path-extension}
  For a DAG ${G_i = (V, E)}$ and candidates nodes ${u, v \in V}$, the
  collapse of ${u \Join v}$ into ${\hat{u}}$ may extend the critical
  path length in ${\hat{G}_i}$: ${\hat{L}_i \ge L_i}$.
\end{obs}

\begin{figure}[ht]
  \centering
  \begin{subfigure}[b]{0.4\columnwidth}{
      \centering
      \includegraphics[height=.6\textwidth]{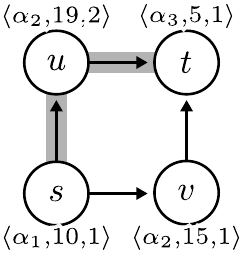}
      \caption{Pre-Collapse ${L_i = 34}$}
      \label{fig:collapse-extend-01}
    }
  \end{subfigure}
  \begin{subfigure}[b]{0.4\columnwidth}{
      \centering
      \includegraphics[height=.6\textwidth]{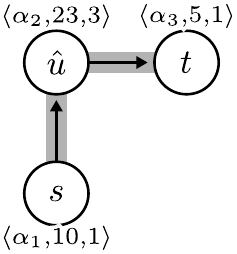}
      \caption{Post-Collapse ${\hat{L}_i = 38}$}
      \label{fig:collapse-extend-02}
    }
  \end{subfigure}
  \caption{Critical Path Extension}
  \label{fig:collapse-extend}
\end{figure}

In contrast to Observation~\ref{obs:critical-path-reduction}, collapse
may extend the critical path length. This can occur when one of the
candidate nodes ${u, v \in V}$ lies on the pre-collapse critical path
and the other does not. In Figure~\ref{fig:collapse-extend}, ${u}$ lies
on the pre-collapse critical path. Collapsing ${\hat{u} \leftarrow u \Join v}$
 increases the critical path length ${\hat{L}_i}$ compared
to ${L_i}$ by ${c_u(\eta_u + \eta_v) - c_u(\eta_u)}$.

\begin{obs}[Workload Decrease]
  \label{obs:workload-decrease}
  For a DAG ${G_i = (V, E)}$ and candidates nodes ${u, v \in V}$, the
  collapse of ${u \Join v}$ into ${\hat{u}}$ reduces the workload, i.e.,
  ${\hat{C}_i \le C_i}$.
\end{obs}

For candidates ${u, v \in V}$, their contribution to the workload of
${C_i}$ is ${c_u(\eta_u) + c_v(\eta_v)}$. The contribution of
${\hat{u} \gets u \Join v}$ to ${\hat{C}_i}$ is
${c_{\hat{u}}(\eta_{\hat{u}}) = c_u(\eta_u + \eta_v)}$. Since,
${c_u(\eta)}$ is a concave function,
${c_u(\eta_u + \eta_v) \le c_u(\eta_u) + c_v(\eta_v)}$ and
${\hat{C}_i \le C_i}$.

\begin{obs}[Collapse Occlusion]
  \label{obs:collapse-occlusion}
  For a DAG ${G_i = (V,E)}$, candidates ${(u, v)}$ and ${(x, y)}$, the
  collapse of ${u \Join v}$ may prevent the collapse of ${x \Join y}$.
\end{obs}

\begin{figure}[ht]
  \centering
  \begin{subfigure}[b]{0.4\columnwidth}{
      \includegraphics[width=\textwidth]{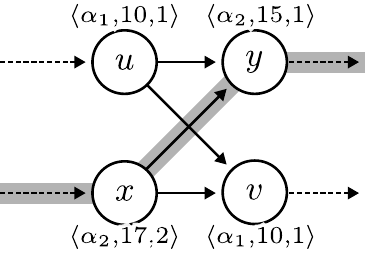}
      \caption{Pre-Collapse}
      \label{fig:collapse-preclude-01}
    }
  \end{subfigure} \quad
  \begin{subfigure}[b]{0.4\columnwidth}{
      \includegraphics[width=\textwidth]{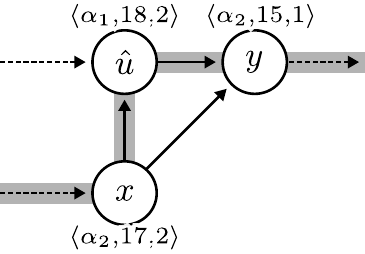}
      \caption{Post-Collapse}
      \label{fig:collapse-preclude-02}
    }
  \end{subfigure}
  \caption{Collapse of ${(u, v)}$ before ${(x, y)}$}
  \label{fig:collapse-preclude}
\end{figure}

Collapsing one candidate ${(u, v)}$ may preclude the collapse of
another. For example, consider Figure~\ref{fig:collapse-preclude}. By
collapsing ${(u, v)}$ the pair ${(x, y)}$ cannot be collapsed -- doing
so would introduce a cycle into the DAG.

~
\begin{table}[ht]
  \centering
  \begin{tabular}{|c|c|c|c|c|c|c|}
    \cline{1-3} \cline{5-7}
    ${C_i}$ & ${L_i}$ & ${m_i}$ & ${u \Join v}$ & ${\hat{C}_i}$ &
    ${\hat{L}_i}$ & ${\hat{m}_i}$\\ 
    52 & 32 & ${\lceil 2.5 \rceil}$ & ${\rightarrow}$ & 50 & 33 &
    ${\lceil 2.42 \rceil}$ \\
    \cline{1-3} \cline{5-7}
  \end{tabular}
  \caption{Collapse of ${u}$ and ${v}$ from Figure~\ref{fig:collapse-preclude}}
  \label{table:collapse-preclude}
\end{table}~

Given a deadline ${D_i = 40}$ the result of collapsing ${(u,v)}$ with
respect to the workload, critical path length, and dedicated cores are
summarized in Table~\ref{table:collapse-preclude}.

\begin{obs}[Alternate Collapse may Decrease ${\hat{m}}$] 
  \label{obs:alternate-collapse}
  For a DAG ${G_i = (V,E)}$, candidates ${(u, v)}$ and ${(x, y)}$, the
  collapse of ${u \Join v}$ which occludes ${x \Join y}$ and resulting
  allocation of cores denoted ${\hat{m}_{(u \Join v)}}$ may be greater
  than the allocation of cores due to collapsing ${x \Join y}$, i.e.,
  ${\hat{m}_{(x \Join y)} < \hat{m}_{(u \Join v)}}$.
\end{obs}

\begin{figure}[ht]
  \centering
  \begin{subfigure}[b]{0.4\columnwidth}{
      \includegraphics[width=\textwidth]{collapse-preclude-01}
      \caption{Pre-Collapse}
      \label{fig:collapse-preclude-01}
    }
  \end{subfigure}
  \begin{subfigure}[b]{0.4\columnwidth}{    
      \includegraphics[width=\textwidth]{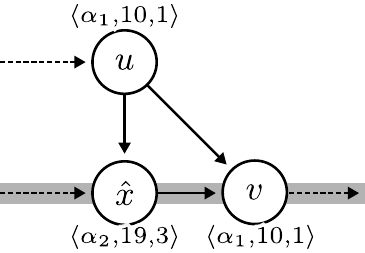}
      \caption{Post-Collapse}
      \label{fig:collapse-preclude-03}
    }
  \end{subfigure}
  \caption{Collapse of ${(x, y)}$ before ${(u, v)}$}
  \label{fig:collapse-alternative}
\end{figure}

Continuing the example, collapsing ${(x,y)}$ precludes the collapse of
${(u, v)}$. Collapsing ${(x, y)}$ instead of ${(u, v)}$ is shown in
Figure~\ref{fig:collapse-alternative}. The impact upon the workload
and critical path length of ${x \Join y}$ differs from that of ${u
  \Join v}$ and ultimately a difference in ${m}$. 

\begin{table}[ht]
  \centering
  \begin{tabular}{|c|c|c|c|c|c|c|}
    \cline{1-3} \cline{5-7}
    ${C_i}$ & ${L_i}$ & ${m_i}$ & ${x \Join y}$ & ${\hat{C}_i}$ &
    ${\hat{L}_i}$ & ${\hat{m}_i}$\\ 
    52 & 32 & ${\lceil 2.5 \rceil}$ & ${\rightarrow}$ & 49 & 29 &
    ${\lceil 1.81 \rceil}$ \\
    \cline{1-3} \cline{5-7}
  \end{tabular}
  \caption{Collapse of ${x}$ and ${y}$ from
    Figure~\ref{fig:collapse-alternative}}
  \label{table:collapse-alternative}
\end{table}

Table~\ref{table:collapse-alternative} illustrates the impact of
ordering of collapse with respect to ${m}$, where collapsing ${x \Join y}$
in place of ${u \Join v}$ yields a smaller number of dedicated cores
${m}$.

\subsection{Beneficial Collapse}

By Observations~\ref{obs:critical-path-reduction}-%
\ref{obs:alternate-collapse}, collapsing any individual candidate may
increase or decrease the number of cores allocated to a task. A
collapse may increase or decrease the critical path length creating an
infeasible task set or introduce a cycle into the graph. This
subsection defines which collapses are \emph{beneficial}.

Beneficial collapse depends on the Definition~\ref{def:improved-m} of
\emph{improving} the allocation of cores. Improving the number of
allocated cores balances the concepts of reducing the number of
cores allocated to a feasible task, avoiding the creation of
an infeasible task, and (possibly) creating feasible
tasks from infeasible ones. 

\begin{definition}[Improved Core Allocation]
  \label{def:improved-m}
  For a given number of cores allocated to a task ${m_i}$,
  ${\hat{m}_i}$ is an improvement upon ${m_i}$ denoted ${\hat{m}_i \ll
    m_i}$ if and only if: 


  \begin{description}
  \item[1.) ${m_i > 0 \Rightarrow 0 < \hat{m}_i \le m_i}$]
  \item[2.) ${m_i \le 0 \Rightarrow \hat{m}_i \ge m_i}$]
  \end{description}
\end{definition}

When ${m_i}$ is greater than zero, an ${\hat{m}_i}$ less than ${m_i}$
and greater than zero is an improvement, reducing the number
of cores allocated to the task. When ${m_i < 0}$, the critical path length
has exceeded the deadline ${L_i > D_i}$. Such a task is not feasible
under the DAG model. For ${m_i}$ less than zero, a ${\hat{m}_i}$ greater than
${m_i}$ is an improvement; an increase over ${m_i}$ may result in a
schedulable task under DAG-OT.

Improvement of ${m_i}$ does not include the ceiling described by
Equation~\ref{eq:m}. This is due to the difference in context of
${m_i}$ under the DAG model compared to DAG-OT. For the DAG model,
${m_i}$ is calculated once and an integer number of cores are assigned
to the task ${\tau_i}$ for schedulability analysis. For the DAG-OT
model, ${m_i}$ is recalculated after every collapse operation. Only
when collapse operations have ceased is the final integer ceiling of
${\hat{m}_i}$ assigned to ${\tau_i}$ for schedulability analysis. The
treatment of ${m_i}$ (and ${\hat{m}_i}$) as a real number rather than
an integer is consistent throughout this work.

Beneficial collapse, given by Definition~\ref{def:beneficial} includes
the improvement of core allocation as one of three conditions. The first
condition maintains the integrity of the analysis, a beneficial
collapse may not introduce a cycle into the graph which the critical
path length calculation depends upon. 

\begin{definition}[Beneficial Collapse]
  \label{def:beneficial}
  For a task ${\tau_i}$, DAG ${G_i = (V, E)}$, and candidate nodes
  ${u, v \in V}$ the collapse of ${u \Join v}$ which results in
  ${\hat{G_i}}$ is \emph{beneficial} if and only if:

  \begin{enumerate}
    \item{${\hat{G}_i}$ contains no cycles}
    \item{${L_i \le D_i \Rightarrow \hat{L}_i \le D_i}$}
    \item{${\hat{m_i} \ll m_i}$}
  \end{enumerate}
\end{definition}

Condition 2 of beneficial collapse definition protects
against collapse increasing the critical path length ${L_i}$ beyond
the deadline ${D_i}$, which would create an unschedulable
task. Protection does not prevent unschedulable tasks becoming
schedulable by collapse, due to the post-collapse critical path length
being bounded by the deadline only if the pre-collapse critical path
length was also less than the deadline.

\subsection{Optimal Collapse}

The primary goal of this work is to improve the schedulability of a
task set by reducing the number of cores reserved for high utilization
tasks. Defining optimality with respect to the number of cores
assigned to a task matches the goal of this work.

\begin{definition}[Optimal Collapse of a Task]
  \label{def:optimal-collapse}
  The \emph{optimal} collapse of a DAG ${G}$ is a DAG ${\hat{G}}$
  with the least positive ${\hat{m}}$ obtainable by collapsing 
  candidates of ${G}$.
\end{definition}

Currently, the complexity class of selecting the optimal set of
candidates to collapse for a single task is unknown and remains an
open problem. Observations~\ref{obs:critical-path-reduction}-%
\ref{obs:alternate-collapse} along with
Definitions~\ref{def:improved-m} and~\ref{def:beneficial} illustrate
the difficulties of identifying candidates that are beneficial to
collapse. The only known method to compute the optimal collapse of a
task requires the exploration of all possible combinations of
candidates. Since there may be ${V^2}$ candidates per task, exploring
all possible combinations has a time complexity of
${\mathbb{O}(2^{V^2})}$. Generating the optimal formulation and
finding an optimal collapse of a task are both potentially intractable
problems and reserved for future work. As a practical alternative,
heuristics for ordering candidates for collapse are
proposed in Section~\ref{sec:selection}.

\section{Collapsing High Utilization Tasks}
\label{sec:dagot-sched}
Due to the intractability of optimal collapse for a task, this work
proposes an intuitive heuristic presented in
Algorithm~\ref{alg:dag-ot-reduce}. It collapses beneficial candidates
(Definition~\ref{def:beneficial}), attempting to reduce the number of
cores allocated to a high utilization task.

\begin{algorithm}[ht]
  \caption{DAG-OT Dedicated Core Reduction
    Algorithm}\label{alg:dag-ot-reduce}
{\footnotesize
  \begin{algorithmic}[1]
    \Procedure{dagot-reduce}{${G_i}$}
        \State ${A \gets}$ \text{\sc{candidates}(${G_i}$)}
            \label{line:candidates}
        \State ${A \gets}$ \text{\sc{order}(${A}$)}
      \While {${|A| \not = 0}$} \label{line:alg-loop}
          \State ${(u,v) \gets }$ \text{\sc{first}(${A}$)} \label{line:choose}
          \State ${A \gets A \setminus (u, v)}$
          \If {\text{\sc{benefit}(${G_i, u, v}$)}}
              \State \text{\sc{collapse}(${G_i, u, v}$)}
          \EndIf
      \EndWhile
    \EndProcedure
  \end{algorithmic}
  }
\end{algorithm}

Reduction begins by identifying the potential candidates for collapse
on Line~\ref{line:candidates}. Candidacy follows
Definition~\ref{def:candidates}. Calculating the complete set of
candidates is of complexity ${\mathbb{O}(V^2)}$. The set of candidates
is prioritized for collapse consideration by \text{\sc{order}}.
Ordering heuristics are proposed in Section~\ref{sec:selection}. Each
proposed heuristic is of equal or lesser computational complexity than
the while loop (and its contents) beginning on Line~\ref{line:alg-loop}.

Only candidates that benefit the task set are collapsed, improving
(Definition~\ref{def:improved-m}) the number of cores allocated to a
task without introducing a cycle into the DAG. The time complexity of
checking for a cycle in ${\hat{G}_i}$ by a depth first search (DFS) is
${\mathbb{O}(V + E)}$. The time complexity of calculating 
${\hat{L}_i}$ of a DAG by topological sort is also ${\mathbb{O}(V +
  E)}$. Determining if the number of allocated cores satisfy
the definition of improvement is an ${\mathbb{O}(1)}$ operation, and
collapse is an ${\mathbb{O}(E)}$ operation. Iterating over
${\mathbb{O}(V^2)}$ possible candidates, time complexity of
Algorithm~\ref{alg:dag-ot-reduce} is ${\mathbb{O}(V^3 + V^2E)}$.

During each iteration of the while loop on Line~\ref{line:alg-loop} of
the \text{\sc{dagot-reduce}} Algorithm~\ref{alg:dag-ot-reduce} the
current state of the DAG ${G_i}$ serves as input and ${\hat{G}_i}$ is
the output. A subsequent iteration of the loop consumes the previous
${\hat{G}_i}$ value as input when considering the next candidate for
collapse.

\section{Candidate Ordering}
\label{sec:selection}

In this work, two heuristics for collapse ordering are proposed:
``greatest benefit'', orders the candidates by descending
workload savings; ``least penalty'', orders candidates by increasing
longest path extension.

\subsection{Greatest Benefit}
For the greatest benefit heuristic, intuition suggests that collapsing
nodes that most reduce the total workload ${C_i}$ will also reduce the
number of cores ${m_i}$ maximally. The difference in 
workload is represented by ${\Delta}$ in Equation~\ref{eq:delta}. 
There is a one time cost to calculate ${\Delta}$ for all candidates in
${A}$ and order the set. This operation is of
${\mathbb{O}(V \lg V)}$ complexity. Employing the greatest benefit
heuristic, Algorithm~\ref{alg:dag-ot-reduce} is then
${\mathbb{O}(V \lg V + V^3 + V^2E) = \mathbb{O}(V^3 +  V^2E)}$
complex.
\begin{equation}
  \label{eq:delta}
  \Delta = c_u(\eta_u) + c_v(\eta_v) - c_u(\eta_u + \eta_v)
\end{equation}

\subsection{Least Penalty}

For the least penalty heuristic, the intuitive reasoning is 
collapsing nodes that impact the critical path length by the
smallest amount (possibly negative) may permit more nodes to be
collapsed overall. Penalties ${\gamma}$ are calculated once by
Equation~\ref{eq:penalty} for every candidate pair. The set of
candidates ${A}$ are ordered by increasing penalty for use in
Algorithm~\ref{alg:dag-ot-reduce}.
\begin{equation}
  \label{eq:penalty}
  \gamma = \hat{L}_i - L_i
\end{equation}

Penalty calculation requires a topological sort for every candidate to
find ${\hat{L}_i}$ with complexity ${\mathbb{O}(V + E)}$, for
${\mathbb{O}(V^2)}$ candidates. Sorting the candidates by penalty is
${\mathbb{O}(V \lg V)}$ complex. Therefore, the initial penalty
ordering complexity is ${\mathbb{O}(V^3 + V^2E + V \lg V)}$. The
complexity of Algorithm~\ref{alg:dag-ot-reduce} utilizing the least
penalty heuristic is then
${\mathbb{O}(V^3 + V^2E + V \lg V + V^3 + V^2E) = \mathbb{O}(V^3 +  V^2E)}$.

Penalty calculations apply to a single DAG ${G_i = (V,E)}$
instance. Collapsing two nodes ${u,v \in V}$ may impact the
critical path length, i.e. ${\hat{L}_i \not = L_i}$. The penalty of
collapse depends on the critical path length, the collapse of ${u
  \Join v}$ may impact the penalty ${\gamma}$ of other
candidates. Penalties are not adjusted after collapsing nodes for the
  least penalty heuristic in favor of maintaining the
${\mathbb{O}(V^3 + V^2E)}$ complexity of
Algorithm~\ref{alg:dag-ot-reduce}.

\section{Collapsing Low Utilization Tasks}
\label{sec:lowutil-collapse}
Previous sections have focused on reducing ${m_i}$ for high
ulitization tasks. Low utilization tasks may also incorporate the
inter-thread cache benefit through collapse. To incorporate the
benefit, a non-preemptive scheduler is required due to \bundle{}'s
lack of preemptive schedulability analysis.

\begin{figure}[ht]
  \captionsetup[subfigure]{position=b}
  \centering
  \begin{subfigure}[b]{0.3\columnwidth}{
      \includegraphics[width=\textwidth]{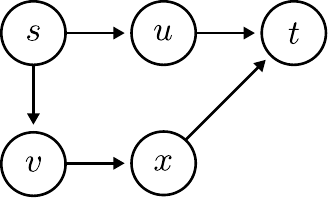}
      \caption{Pre-Serializing}
      \label{fig:serialize-01}
    }
  \end{subfigure} \quad
  \begin{subfigure}[b]{0.5\columnwidth}{
      \includegraphics[width=\textwidth]{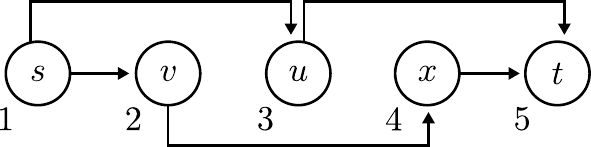}
      \caption{Post-Serializing}
      \label{fig:serialize-02}
    }
  \end{subfigure}
  \caption{Serializing a Task ${\tau_i}$}
  \label{fig:serialize}
\end{figure}

A low utilization DAG task ${\tau_i \in \tau_{low}}$ requires no more than one core
${m_i = 1}$ to meet all deadlines. Therefore, ${\tau_i}$ may be
\emph{serialized}. To serialize ${\tau_i}$ a topological
sort of ${G_i}$ is performed and nodes are executed on the single
processor in sort order. Figure~\ref{fig:serialize} illustrates the
serialization of a task ${\tau_i}$.

Before a low utilization task is serialized all beneficial
(Definition~\ref{def:beneficial}) candidates ${u, v \in V_i}$
collapsed. For a serialized task ${\tau_i}$, the workload bounds the
critical path length ${C_i \ge L_i}$. A serialized task is infeasible
if ${C_i > D_i}$. Since the workload is only reduced by collapse,
collapse preceding serialization cannot convert a feasible task into
an infeasible one.

Similar to high utilization tasks, the complexity of serializing low
utilization tasks depends on the number of candidates
${\mathbb{O}(V^2)}$, a DFS to check for cycles ${\mathbb{O}(V + E)}$,
and a topological sort to order execution ${\mathbb{O}(V + E)}$. The
total complexity of the operation is
${\mathbb{O}(V^2 \cdot (V + E) + (V + E)) = \mathbb{O}(V^3 + V^2E)}$.

Another concern shared with high utilization tasks is the order of
collapse. For simplicity, collapse ordering is defined for the entire
task set and shared between high and low utilization tasks. Whichever
heuristic is selected for high utilization tasks is also selected for
low utilization tasks for all tasks ${\tau_i \in \tau}$.

Every collapsed and serialized low utilization task
${\tau_i \in \tau_{low}}$ is scheduled non-preemptively, lest the
inter-thread cache benefit of scheduling individual threads of nodes
via \bundle{} be lost. To preserve the benefit of collapse and
\bundle{}, low utilization jobs are scheduled by non-preemptive EDF.

Each low utilization task ${\tau_i \in \tau_{low}}$ is assigned to
exactly one of the ${m_{low}}$ cores by the
\emph{Worst-Fit}~\cite{baruah2013partitioned}\footnote{Any
  non-preemptive EDF schedulability test based task assignment to 
  cores could be chosen.} heuristic.  Worst-Fit assigns each task
${\tau_i \in \tau_{low}}$ to a per-core task set on a core ${m_k}$
when including ${\tau_i}$ will not create an infeasible per-core task
set determined by~\cite{jeffay1991}. Once assigned, jobs of ${\tau_i}$
will execute only upon its assigned processor.

\section{Scheduling and Schedulability Analysis}
\label{sec:sched-analysis}

Federated scheduling and schedulability
  analysis~\cite{li2014analysis} of parallel DAG task sets may be
  considered for separate task sets: high and low utilization tasks. For
  any high utilization task ${\tau_i \in \tau_{high}}$, any greedy
  non-preemptive scheduler may be used to select which node to execute
  upon one of the ${m_i}$ cores dedicated to the task. For low
  utilization tasks ${\tau_{low}}$, a preemptive or non-preemptive
  multi-core scheduling algorithm may be used to execute
  nodes upon the ${m_{low}}$ cores.

Schedulability analysis of high utilization tasks follows from
  two conditions. The first is the requirement that a task ${\tau_i
    \in \tau_{high}}$ must have a critical path length less than its
  deadline ${L_i < D_i}$. The second is that ${\tau_i}$ has
  ${m_i}$ cores allocated as calculated by Equation~\ref{eq:m}. If
  there are an insufficient number of cores in the system to satisfy
  all high utilization tasks i.e. ${m < m_{high} = \sum_{\tau_i \in
  \tau_{high}} m_i}$, the task set is unschedulable.

In this work, low utilization tasks are scheduled by partitioned
  EDF to the ${m_{low} = m - m_{high}}$ cores. For DAG tasks, that may
  be preemptive or non-preemptive EDF. For DAG-OT tasks, it must be
  non-preemptive EDF or the benefits of \bundle{} scheduling cannot be
  guaranteed. In partitioned EDF, tasks are assigned to cores. In the
  preemptive and non-preemptive setting, tasks are assigned to cores
  by the \emph{Worst-Fit}~\cite{baruah2013partitioned}
  heuristic. Under Worst-Fit partitioning, a task will not be assigned
  to a core if assigning it would violate the uniprocessor scheduling
  test. The uniprocessor non-preemptive schedulability
  test is taken from~\cite{jeffay1991} and the preemptive
  schedulability test from~\cite{baruah:1990}.

In summary, a taskset is deemed schedulable if all high
  and low utilization tasks are schedulable. For high utilization
  tasks, there must be a sufficient number of ${m_{high}}$ cores and
  all critical paths are less than deadlines. For low utilization
  tasks, all tasks must be partitioned on the ${m_{low}}$ cores
  according to Worst-Fit without violating the appropriate per-core
  schedulability test~\cite{jeffay1991} or~\cite{baruah:1990}.

\section{Evaluation}
\label{sec:evaluation}

Evaluation of the approach proposed in this work focuses on two
metrics: \emph{schedulability ratios} and the \emph{reduction of
  dedicated cores} to 
high utilization tasks. No existing approach to federated scheduling
tasks incorporating the positive impact of instruction
caches is (currently) known. To illustrate the potential of
inter-thread cache benefits to DAG tasks under federated
scheduling~\cite{li2014analysis}, high utilization tasks are scheduled
by any non-preemptive work-conserving algorithm on 
the cores dedicated to the individual tasks. Low utilization tasks are
assigned to cores by the Worst-Fit~\cite{baruah2013partitioned}
partitioning algorithm and scheduled by non-preemptive EDF. In
addition to the non-preemptive EDF scheduling of low utilization tasks, a
comparison to federated scheduling using preemptive EDF of low
utilization tasks is made. For preemptive EDF, it is assumed that
preemptions have no preemption cost. As the proposed approach uses
non-preemptive scheduling for scheduling low utilzation tasks, this
assumption only benefits the previous federated scheduling schemes
which require preemption.

To permit larger scale testing, if any schedulability test for a task
set takes more than 10 minutes to complete, then that task set is deemed
unschedulable for the given test. For fairness across the heuristics,
such a task set is deemed unschedulable for all heuristic collapse
methods.

The existing schedulability analysis approaches are compared to
collapse by \text{\sc{dagot-reduce}} using the proposed heuristics.
The proposed heuristics are also compared against an ``arbitrary''
(random) ordering to highlight each heuristic's impact.
Table~\ref{table:comparisons} summarizes the existing and proposed
approaches used in the evaluation along with their notation. The
approaches are enumerated by their inclusion of collapse and their use
of non-preemptive EDF (EDF-NP) or preemptive EDF (EDF-P) for low
utilization tasks.

\begin{table}[ht]
  \centering
  \begin{tabular}{|r|l|c|}
    \hline
    Approach & EDF-NP & EDF-P \\
    \hline
    Baseline (No Collapse) & B-NP & B-P \\
    Collapse Arbitrary & OT-A & ${\varnothing}$ \\
    Collapse Greatest Benefit & OT-G & ${\varnothing}$ \\
    Collapse Least Penalty & OT-L & ${\varnothing}$ \\
    \hline
  \end{tabular}
  \caption{Federated Schedulability Tests to Compare}
  \label{table:comparisons}
\end{table}

Synthetic task sets are provided to each of the schedulability
tests. Generation of the synthetic DAG tasks takes the form of a
pipeline, where individual tasks are synthesized and then combined to
make task sets. To allow a comparison to be made between the baseline
and collapsed tasks, tasks are generated for the baseline first and
then collapsed. \text{\sc{dagot-reduce}} modifies the structure of DAG
tasks, as well as the critical path length and total demand. Due
collapse related changes, tasks that were trivially infeasible (i.e., ${L_i >
  D_i}$) may become feasible. As such, existing
approaches~\cite{Ueter:2018} to task set generation which exclude
trivially infeasible tasks are unsuitable for this work. 

\begin{figure}[ht]
  \centering
  \includegraphics{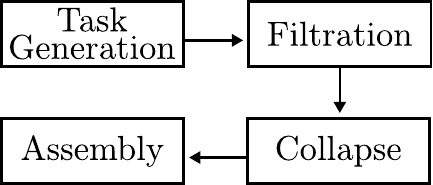}
  \caption{Task Set Generation Pipeline}
  \label{fig:eval-pipeline-01}
\end{figure}

  Figure~\ref{fig:eval-pipeline-01} describes the pipeline in coarsest
  detail. Individual tasks are generated, and filtered. The full
  details of the task set generation pipeline can be found in the
  Appendix, the framework is available for download
  on github~\cite{libsched:2020, evaluation:2019}. 

  Summarily, tasks are generated by creating a representative set of 90 DAG
  task graphs of \{16, 32, 64\} nodes, with a variable edge probability between
  each pair of nodes. For each graph structure, nine graphs are generated
  by parameterizing the number of executable objects \{4,8,16\} and their
  growth factors \{0.2, 0.6, 1.0\}. For each task, six new tasks are
  generated with deadlines calculated using a target utilization
  \{0.25, 0.50, 2.0, 4.0, 8.0, 16.0\}. In total, 4,860 tasks are generated.

  Filtration of the 4,860 tasks removes only those tasks which are
  trivially infeasible ${(L_i > D_i)}$ for the baseline DAG task and
  for all collapsed DAG-OT tasks. It should be noted that any
  post-collapse DAG-OT task ${\hat{\tau}_i}$ which is trivially
  infeasible could not have originated from a pre-collapse trivially
  infeasible DAG task ${\tau_i}$. The filtered tasks are then
  duplicated, once per collapse ordering, before being assembled into
  task sets.

Table~\ref{table:eval-assembly} enumerates the total number of task
sets created by target task set utilization ${U}$, cores of the
architecture ${c}$, and number of task sets assembled per utilization
and core specification ${N}$. The total reflects the total number of
DAG task sets assembled, it does not reflect the equivalent DAG-OT
task sets (resulting from collapse by each of the heuristics).

\begin{table}[ht]
  \centering
  \begin{tabular}{|c|c|}
    \hline
    Parameter & Range \\
    \hline
    ${U}$ & ${\{0.5, 1, 2, 4, 8, 12, 16, 20, 24, 28, 32, 36\}}$ \\
    ${c}$ & ${\{4, 8, 12, 16, 20, 24, 28, 32\}}$ \\
    ${N}$ & 1000 \\
    \hline
    \hline
    Total & 
    {${ N \cdot | c | \cdot |U|  = 96,000}$} \\
    \hline
  \end{tabular}
  \caption{Task Set Assembly Parameters}
  \label{table:eval-assembly}
\end{table}

\subsection{Evaluation Metrics}

A schedulability ratio is calculated for each of the schedulability
tests. For the DAG-OT schedulability tests, the number of cores saved
${m_{i,saved}}$ per task ${\tau_i}$ is calculated by
Equation~\ref{eq:cores-saved} where pre-collapse ${m_{i,high}}$ comes
from Equation~\ref{eq:m} and ${\hat{m}_{i,high}}$ after
Algorithm~\ref{alg:dag-ot-reduce} has terminated.
\begin{equation}
  \label{eq:cores-saved}
  m_{i,saved} = m_{i,high} - \hat{m}_{i,high}
\end{equation}

For a task set ${\tau}$, the change in number of cores allocated
to high utilization tasks is given by Equation~\ref{eq:cores-change-avg}.
\begin{equation}
  \label{eq:cores-change-avg}
  \Delta_m = \sum_{\tau_i \in \tau} m_{i,saved}
\end{equation}

For a DAG-OT task ${\hat{\tau}_i}$ collapsed from a DAG
task ${\tau_i}$, the workload reduction and critical path length
extension of collapse are computed by Equations
\ref{eq:workload-change-avg} and \ref{eq:critical-path-change-avg} respectively.
\begin{align}
  \Delta_{C} &= C_i - \hat{C}_i
  \label{eq:workload-change-avg} \\
  \Delta_{L} &= \hat{L}_i - L_i
  \label{eq:critical-path-change-avg}
\end{align}
\subsection{Results}

\begin{wrapfigure}[10]{l}{0.49\columnwidth}
  \includegraphics[width=0.49\columnwidth]{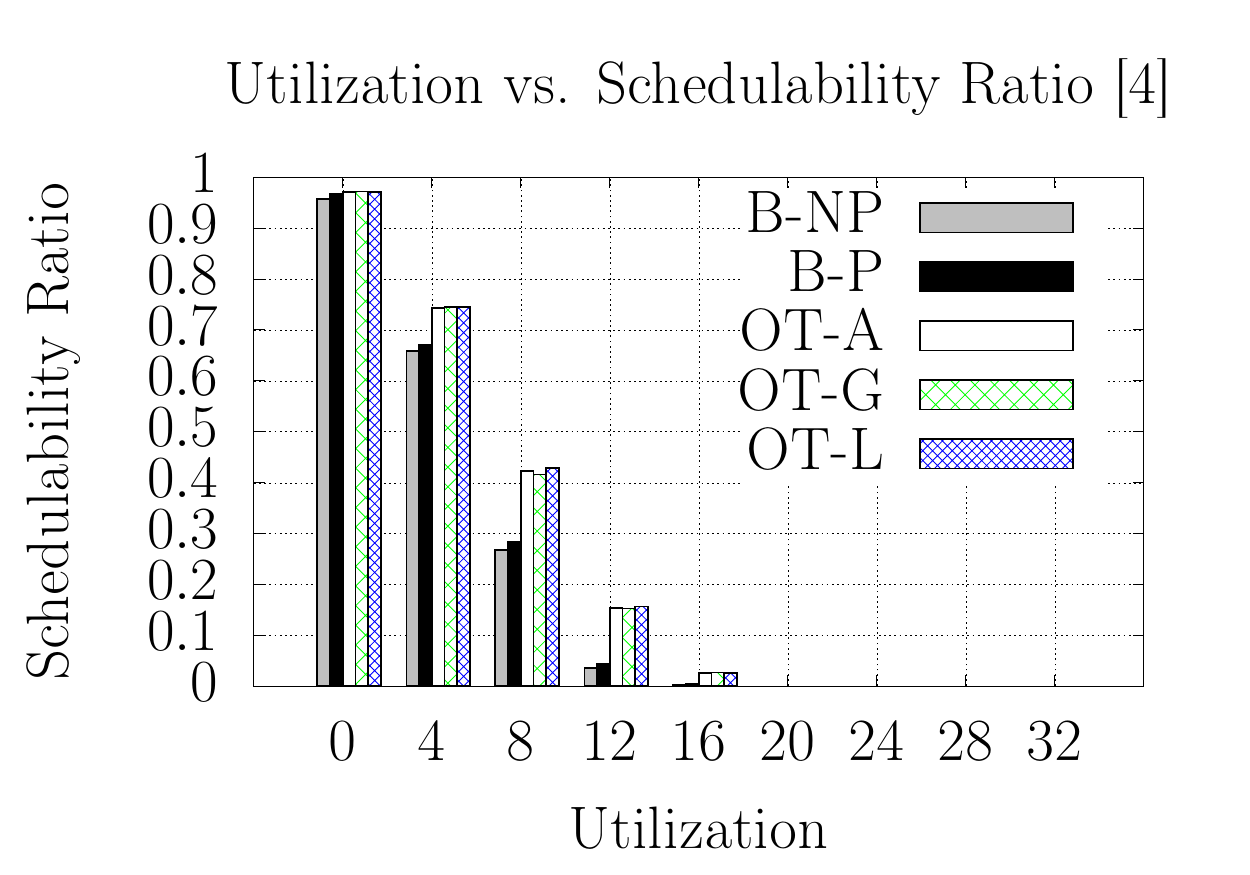}
  \caption{Mean Schedulability}
  \label{fig:summary-01}
\end{wrapfigure}

Figure~\ref{fig:summary-01} summarizes the schedulability results. In
the title '4' indicates the utilization interval the column
summarizes. For the histograms labeled '0', the utilization
schedulability ratio is for task sets from with utilization ${[0,
4)}$. The height of the bar is the average schedulability ratio
over the interval. From this summary data, it is clear that collapse
improves the schedulability of federated scheduled DAG
tasks where collapsed task sets have a higher schedulability ratios.

Furthermore the gains in schedulability from collapsing outweigh any
deleterious effects of the non-preemptive scheduling requirement for
DAG-OT. This can be observed through the higher schedulability ratios for
collapsed task sets compared to the uncollapsed
\emph{fully preemptive} low utilization task sets of B-P. The
fully preemptive scheduler incurs no penalty for preemptions between
low utilization tasks. 

Requiring consideration for trivially infeasible tasks where the
critical path length exceeds the deadline (${L_i > D_i}$), 
constraints found in other works for task set formulation are
prohibited. For example, in~\cite{saifullah:2014} the minimum period
for an arbitrary period task ${\tau_i}$ is ${T_i = L_i + 2
  \frac{C_i}{m}}$. Due to implicit deadline ${T_i = D_i}$ tasks, no 
arbitrary period task in~\cite{saifullah:2014} will require more than
${\frac{m}{2}}$ cores. In this work, no such constraint is possible
resulting in tasks requiring up to ${V_i}$ cores. Consequentially,
higher utilization task sets assembled from tasks with core
allocations upper bounded by the number of nodes in a task are more
likely to be unschedulable. Thus, schedulability ratios for
utilizations over twenty are near zero.

\begin{figure}[ht]
  \centering
  \begin{subfigure}[b]{0.49\columnwidth}{
      \includegraphics[width=\textwidth]
                      {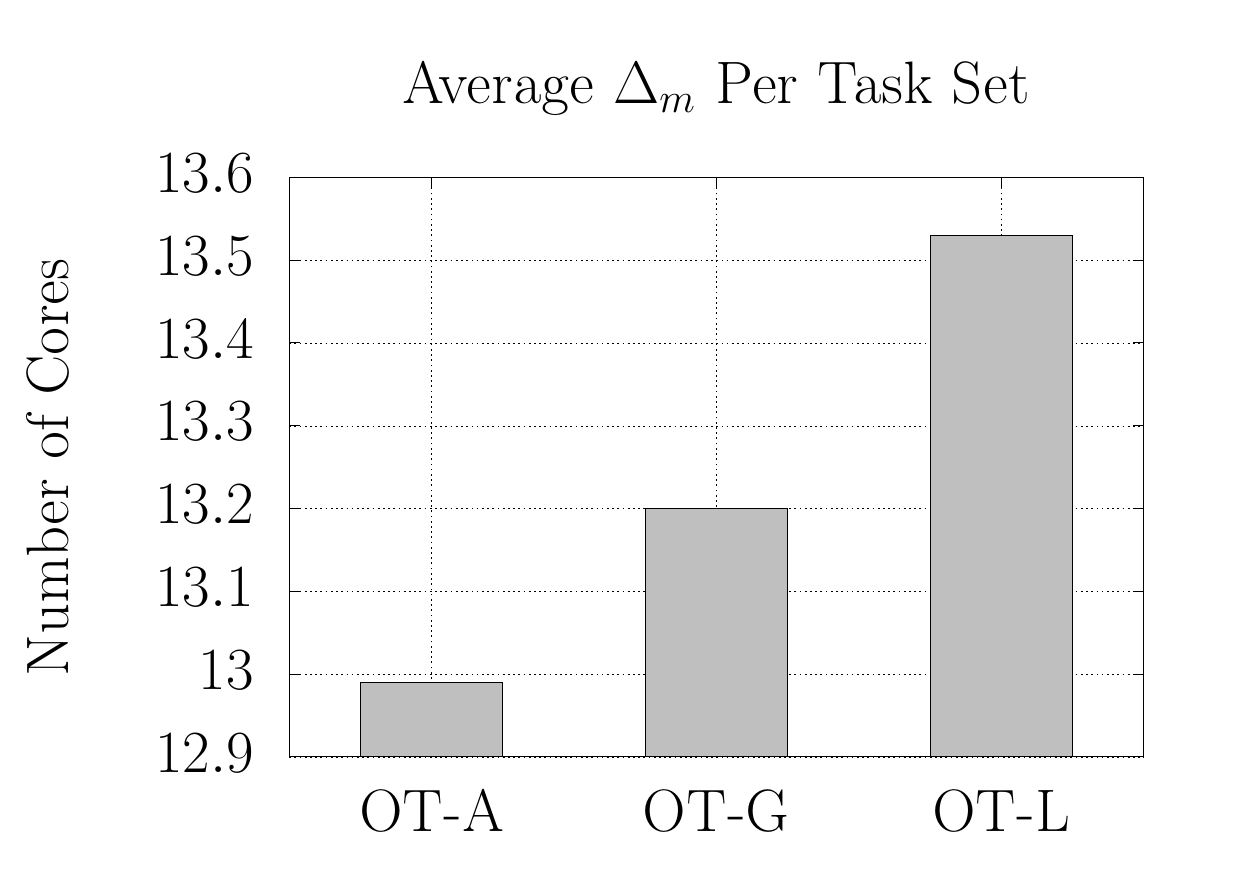}
      \caption{Mean Core Savings}
      \label{fig:summary-02-a}
    }
  \end{subfigure}
  \begin{subfigure}[b]{0.49\columnwidth}{
      \includegraphics[width=\textwidth]
                      {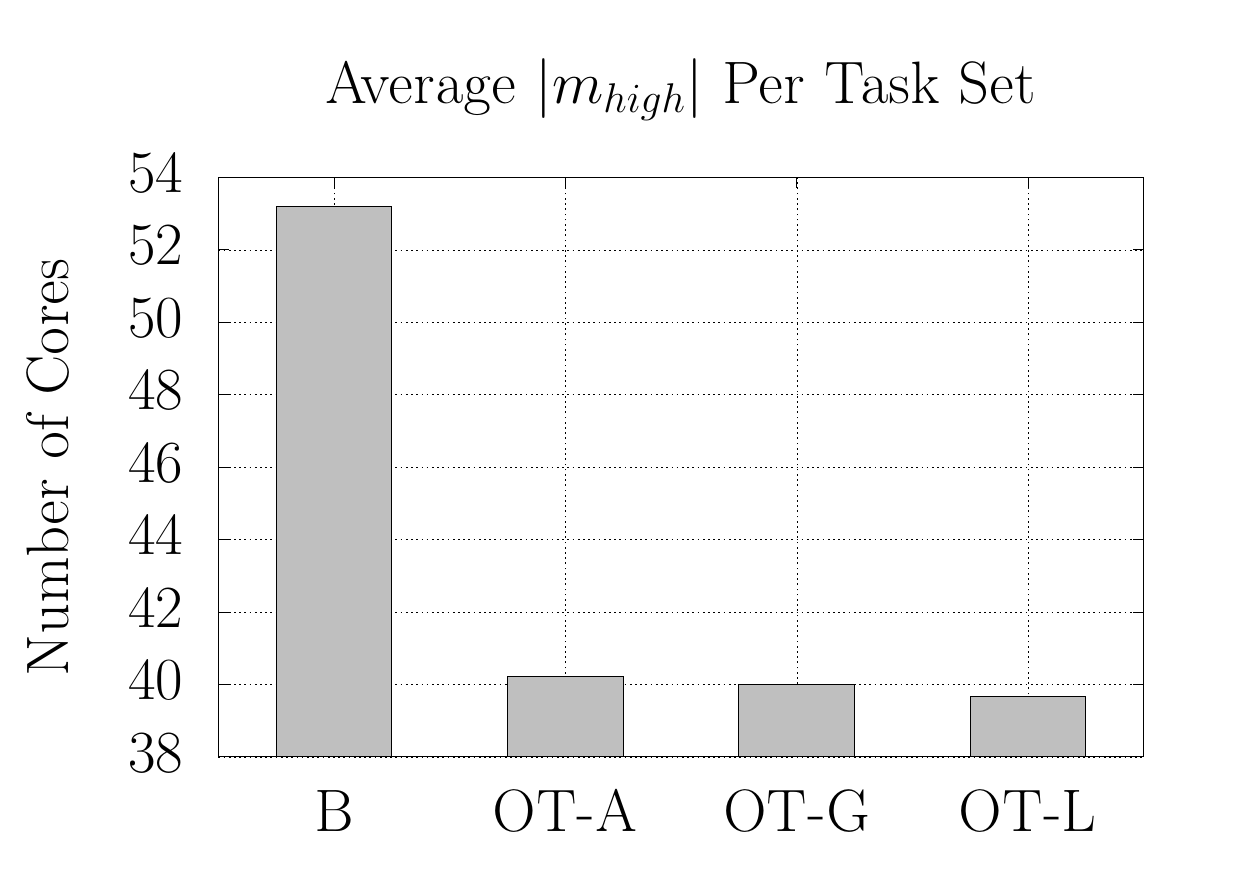}
      \caption{Mean Cores}
      \label{fig:summary-02-b}
    }
  \end{subfigure}
  \caption{Mean Cores and Savings}
  \label{fig:summary-02}
\end{figure}

\begin{figure}[ht]
  \centering
  \begin{subfigure}[b]{0.49\columnwidth}{
      \includegraphics[width=\textwidth]
                      {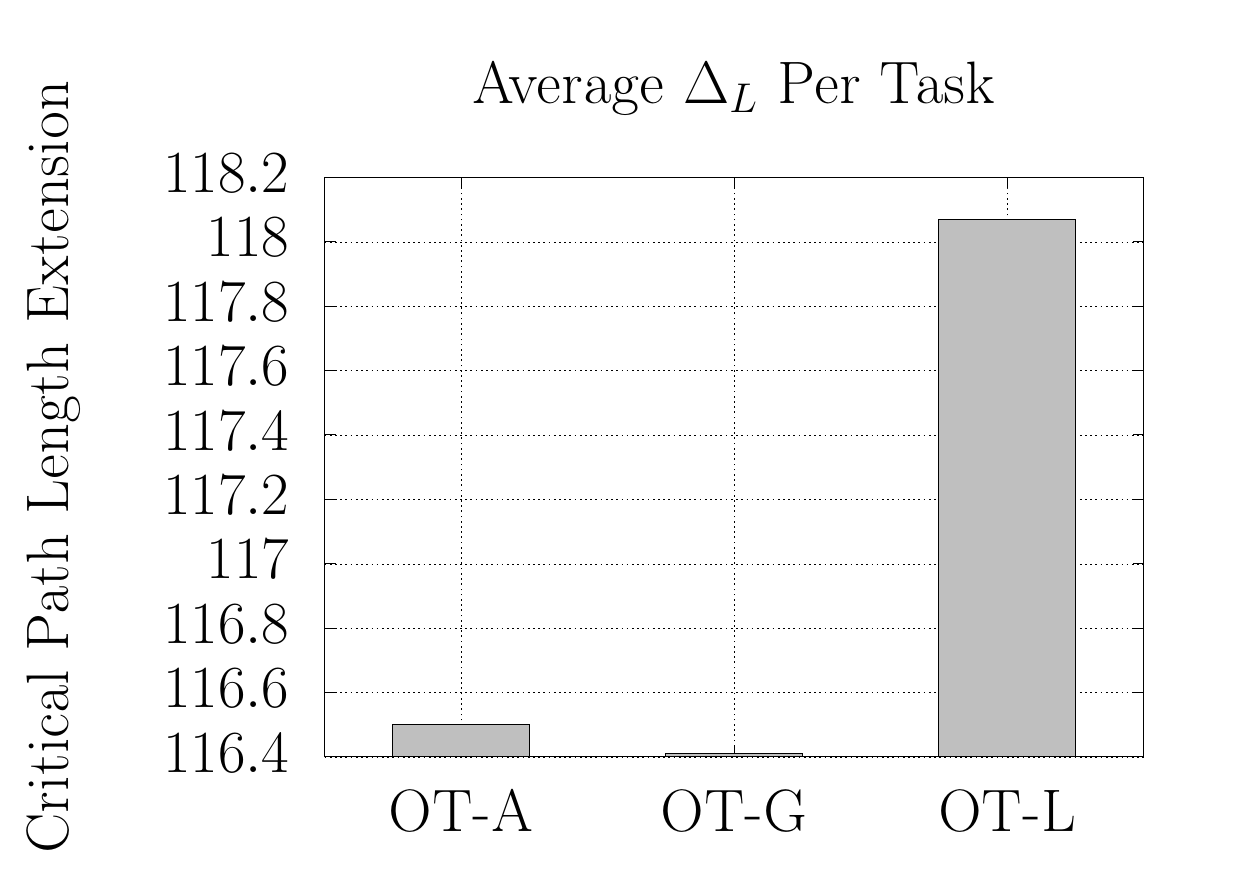}
      \caption{${\bar{\Delta}_L}$}
      \label{fig:summary-03-a}
    }
  \end{subfigure}
  \begin{subfigure}[b]{0.49\columnwidth}{
      \includegraphics[width=\textwidth]
                      {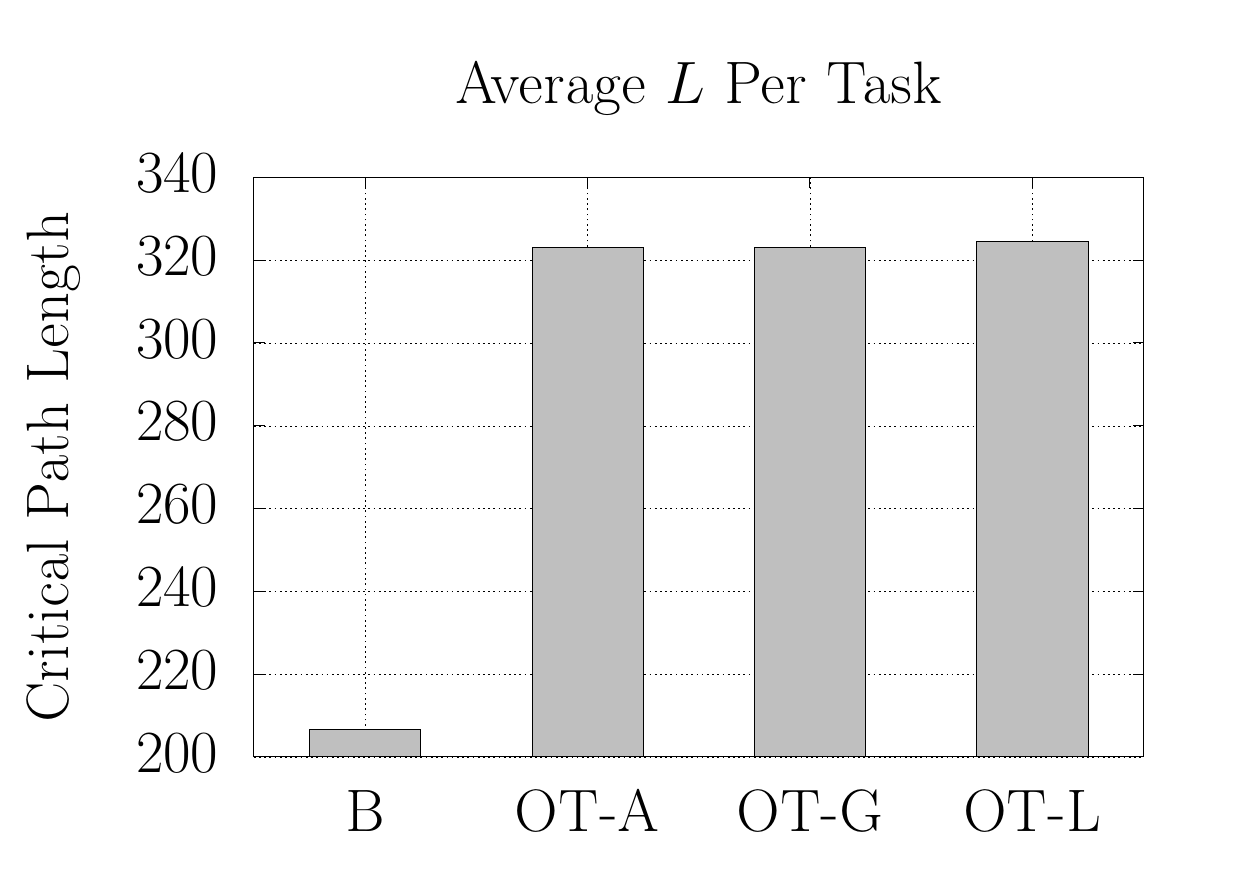}
      \caption{${\bar{L}}$}
      \label{fig:summary-03-b}
    }
  \end{subfigure}
  \caption{Mean Critical Path Lengths and Extensions}
  \label{fig:summary-03}
\end{figure}

\begin{figure}[ht]
  \centering
  \begin{subfigure}[b]{0.49\columnwidth}{
      \includegraphics[width=\textwidth]
                      {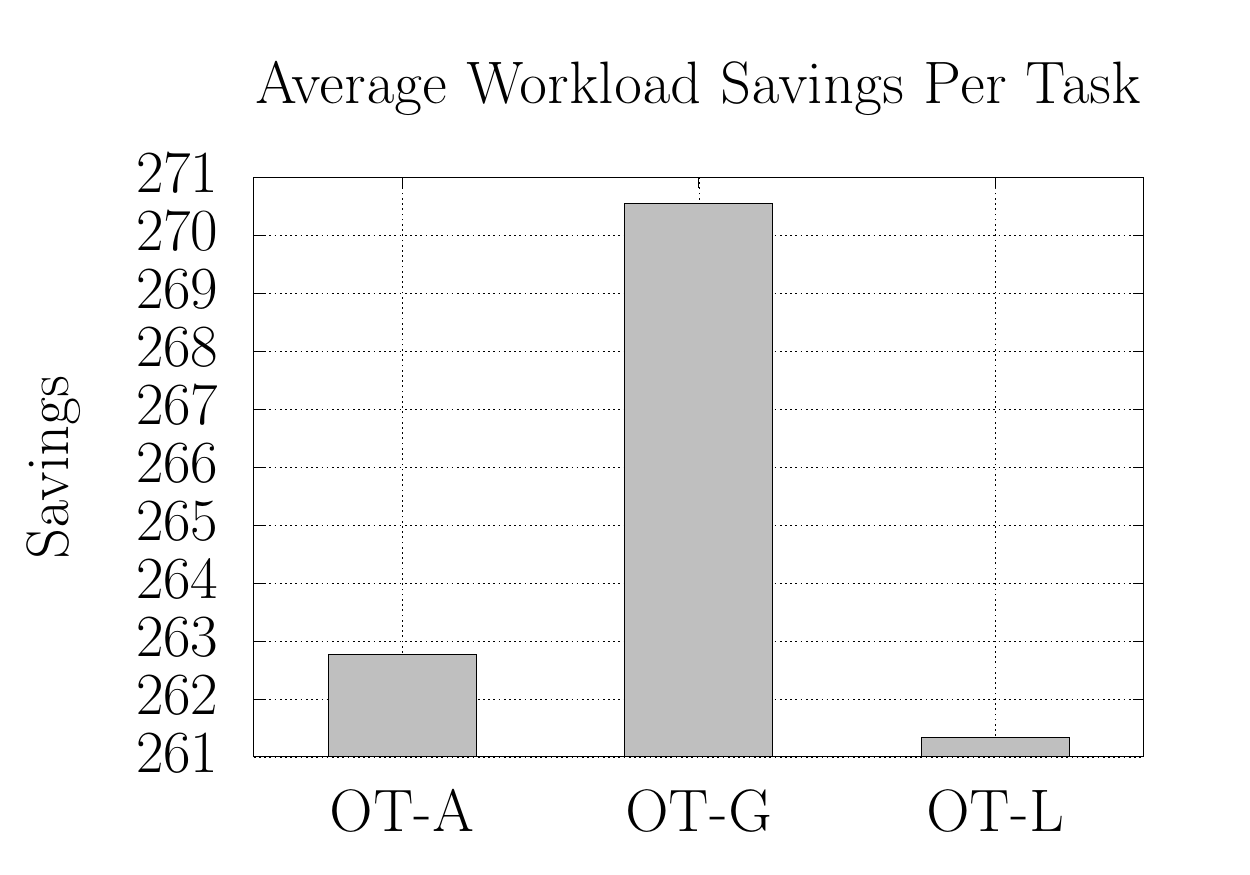}
      \caption{${\bar{\Delta}_C}$}
      \label{fig:summary-04-a}
    }
  \end{subfigure}
  \begin{subfigure}[b]{0.49\columnwidth}{
      \includegraphics[width=\textwidth]
                      {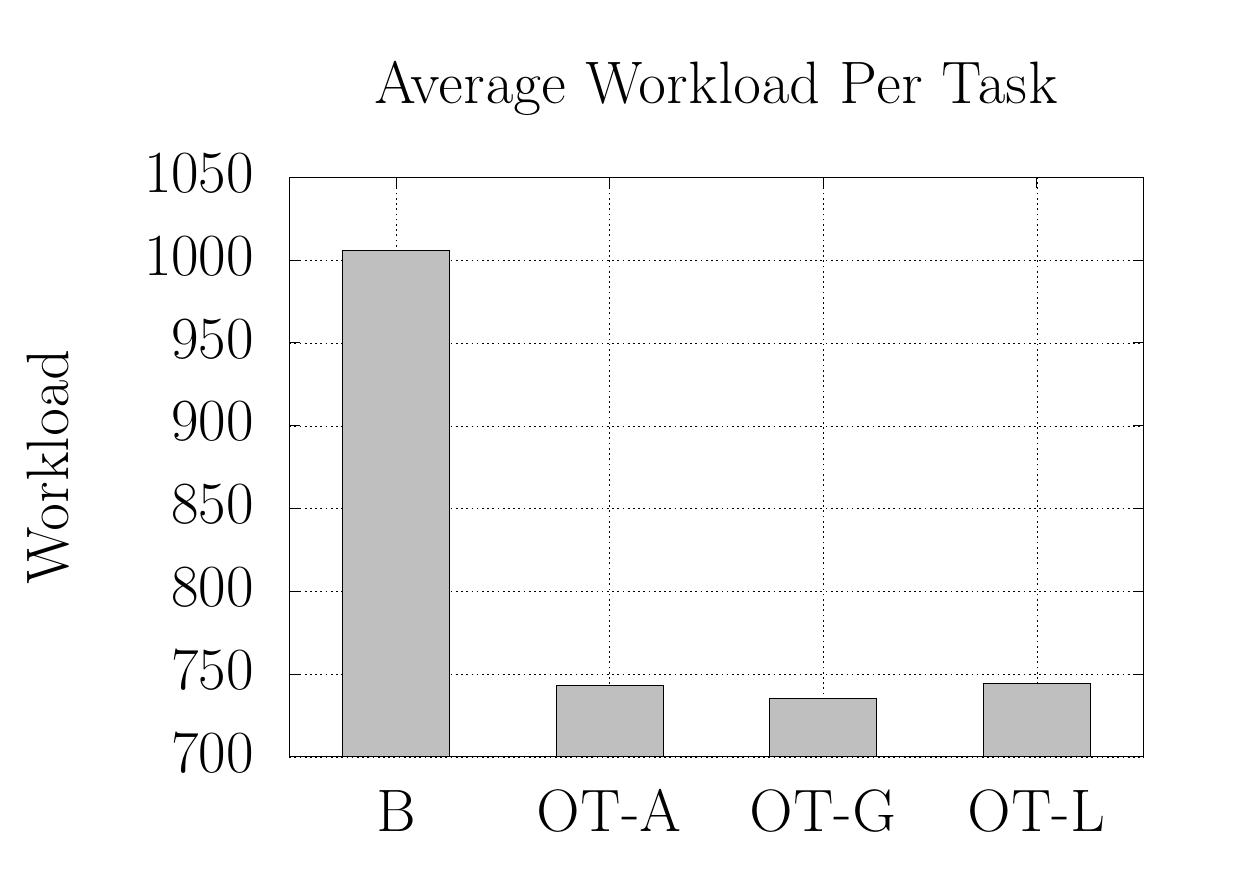}
      \caption{${\bar{C}}$}
      \label{fig:summary-04-b}
    }
  \end{subfigure}
  \caption{Mean Workloads and Savings}
  \label{fig:summary-04}
\end{figure}

It is unclear from Figure~\ref{fig:summary-01} which of the collapse
heuristics is the most desirable. For different utilization intervals,
the heuristic with highest schedulability ratio may
differ. Figures~\ref{fig:summary-02}-\ref{fig:summary-04} 
present the impact of collapse for metrics other schedulability. Both
the preemptive (B-P) and non-preemptive (B-NP) un-collapsed baseline
methods share the same task sets and therefor the same metrics. The
label B represents both un-collapsed baselines in the
figures. Figure~\ref{fig:summary-02} focuses on the central purpose of 
collapse: to reduce the number of cores assigned to high utilization
tasks. The least penalty heuristic (OT-L) performs better than
greatest benefit (OT-G). With arbitrary collapse ordering (OT-A)
performing below the heuristics. For these task sets, the OT-L heuristic
provides an approximately 20\% reduction in dedicated cores, greater
than arbitrary or OT-G ordering for collapse. 

The least penalty (OT-L) heuristic seeks to collapse those nodes with
the smallest increase to the critical path length before
others. Surprisingly, Figure~\ref{fig:summary-03} shows the least
penalty ordering of collapse may not have the intended effect. For
OT-L, the average critical path length is greater than greatest
benefit (OT-G) or arbitrary ordering (OT-A); although it remains
within 2 percent of OT-G and OT-A.

Figure~\ref{fig:summary-04} illustrates the benefits of
collapse to the workload for all orderings, with OT-G providing the
greatest workload reduction of 27 percent; a 5 percent improvement
over OT-L. 

From the results of Figure~\ref{fig:summary-02},
\ref{fig:summary-03}, and \ref{fig:summary-04} the OT-G
heuristic performs similarly to OT-L in terms of 
core savings, the primary purpose of collapse. However, OT-G
out-performs OT-L in both workload savings and critical path length
extension. This is due to the nature of critical path length
extension in comparison to workload savings. With each collapse, there
is potential for the critical path to shift from one set of nodes to
another. If the critical path length shifts, the initial least penalty
ordering may no longer be in descending critical path length
extension order. Workload savings are not affected when
the critical path shifts; thus greatest benefit provides more
consistent behavior and overall better performance.

\section{Feasibility Study}
\label{sec:poc}

To compliment the synthetic results, a feasibility study (FS) was
developed using the TacleBench~\cite{taclebench} benchmarks executing
on Raspberry Pi 3 devices. The purpose of the FS is to verify
the potential benefit of collapse and the concave growth of WCETO
values from \bundle{} to parallel DAG tasks.

Existing parallel programming environments such as OpenMP and
  CilkPlus lack features for controlling thread scheduling and
  management with respect to cache behavior. Additionally, \bundle{}'s
  analysis~\cite{tessler:2018} is limited to MIPS processors. The
  \bundle{} scheduling algorithm presented in~\cite{tessler:2018}
  requires the use of a MIPS simulator. Lacking an ideal environment
  and platform for the FS, Raspberry Pi
  3 devices were selected for their cost and limited hardware
  components. 

\bundle{}'s analysis and scheduling algorithm have not been
  implemented for ARM processors or Raspberry Pi 3 devices.
  Existing WCET tools~\cite{aiT:2020} for ARM processors do 
  not provide cache analysis. Lacking a WCET tool that provides
  adequate cache analysis, representative WCET and growth factor
  values are calculated based upon observed execution times.
  A representative WCET of a TalceBench benchmark is the
  maximum number of cycles from the observed worst-case response time
  (WCRT) from the set of multiple distinct executions upon a Raspberry
  Pi 3 device. Representative growth factors are calculated from the
  WCRT of ${\eta}$ threads executed sequentially upon a single core --
  e.g. the benefit of three threads scheduled by \bundle{} is
  estimated by executing three threads one after another on one
  core. These representative values are not reliable since they are
  based upon observations rather than analysis and could 
  not be used in an environment where deadlines must be kept.

In \bundle{} scheduling, executable objects are divided into
  conflict free regions. Thread execution is coordinated by region, 
  thereby maximizing the inter-thread cache benefit. In comparison,
  sequential scheduling of threads, where each thread executes from
  the first instruction to the last, has (potentially) fewer
  inter-thread cache benefits~\cite{tessler:2016,tessler:2018}. Thus,
  representative growth factor values from
  sequential execution produces greater (worse) values than \bundle{}
  analysis would provide. This over-estimation biases the results
  against the proposed approach.

\begin{figure}[ht]
  \centering
  \includegraphics[width=0.25\textwidth]{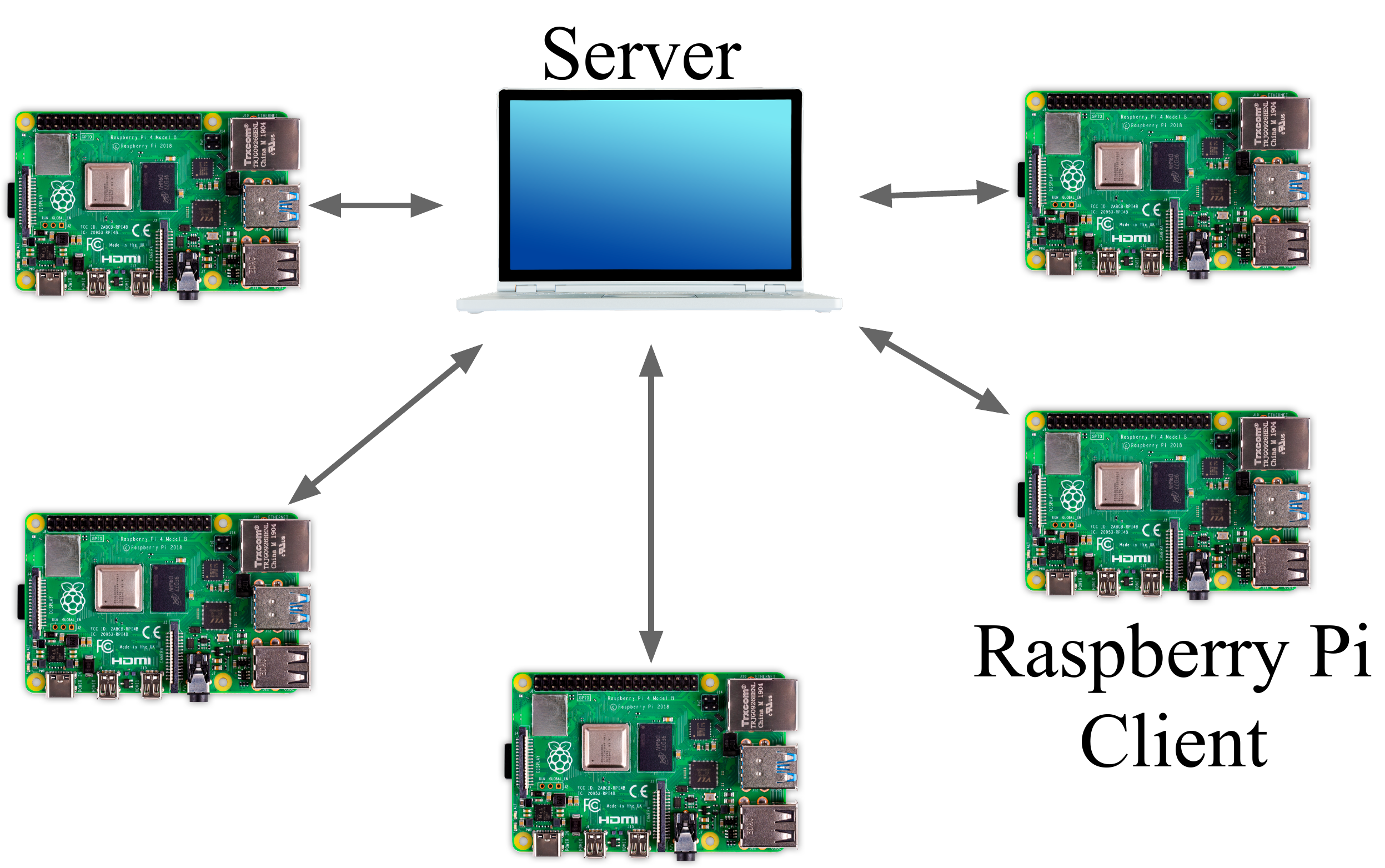}
  \caption{Experiment Architecture}
  \label{fig:exp_architecture}
\end{figure}

The FS is comprised of a process server running on a general
  purpose computer and Raspberry Pi 3 devices, where each Pi device
  represents a single core. The process server schedules nodes of
  the DAG-OT task non-preemptively in a greedy fashion. Execution of a
  node is the sequential execution of a benchmark upon one of the
  Raspberry Pi's representing a core. 
  Figure~\ref{fig:exp_architecture} illustrates the computing  
  platform architecture available for download~\cite{Modekurthy:2019}.

Raspberry Pi 3 devices contain a Samsung four-core ARM Cortex
A53 processor with 32KB of L1 instruction cache, 32KB of
L1 data cache, and 512KB of L2 unified data and
  instruction cache~\cite{ARMManual:2020}. All 
Raspberry Pi 3 devices utilize the Raspbian operating
system running Linux kernel 4.18. To minimize interference from the
operating system as well other processes one of the four cores is
reserved exclusively for the execution of benchmarks.

The sequential execution of benchmarks may benefit from data
values persisting in the cache, thereby decreasing growth factors
and biasing the results towards the proposed approach.
To address the potential bias an attempt at flushing the data
  cache is performed between benchmark
  executions. Algorithm~\ref{alg:POC-main} illustrates the method of
  sequential thread scheduling including data cache flushing. Note,
  the flush may be incomplete due to the pseudo-random L1 and L2
  replacement policy~\cite{ARMManual:2020}.

Each main function from the TacleBench suite is
  modified according to Algorithm~\ref{alg:POC-main}, where ${m}$ is
  a command line argument for the number of threads to execute.
  The \text{\sc{read\_cycles}()} function reads
  the current cycle count of the processor. The
  \text{\sc{old\_main}()} function is the TacleBench provided main
  function. The \text{\sc{clear\_data\_cache}()} function reads 512KB
  (the size of the L2 cache) of allocated memory in an attempt
  to flush the data cache.

\begin{algorithm}[H]
  \caption{Sequential Thread Execution}\label{alg:POC-main}
  {\footnotesize
  \begin{algorithmic}[1]
    \Procedure{main}{m}
        \State ${total \gets 0}$
        \For{$i \gets 1$ \text{to} $m$}
            \State ${pre \gets}$ \text{\sc{read\_cycles}()}
            \State \text{\sc{old\_main}()}
            \State ${post \gets}$ \text{\sc{read\_cycles}()}
            \State ${total \gets total + post - pre}$
            \State \text{\sc{clear\_data\_cache}()}
        \EndFor
    \EndProcedure
  \end{algorithmic}
  }
\end{algorithm}

Algorithm~\ref{alg:POC-main} executes a benchmark once per
  iteration of the for loop, completing ${m}$ threads
  sequentially. After each benchmark execution, the response time  
  is measured by taking the difference of the processor cycle count
  before and after execution. The difference is added to
  the $total$ to compute the total number of cycles required to
  execute $m$ threads of a benchmark. After every execution of a
  benchmark, \text{\sc{clear\_data\_cache}()} performs a partial flush
  of the data cache. The goal is to limit the contribution to the
  inter-thread cache benefit through the data cache.

The FS uses the sample input data from the TacleBench suite for
  every execution. Using the same input for each thread scheduled
  sequentially is an approximation of the inter-thread cache benefit
  \bundle{} scheduling would produce. Under these circumstances, the
  sequential execution provides a lower bound on the inter-thread
  cache benefit producing greater (worse) growth factors than
  \bundle{} scheduling would.

Each of the 42 benchmarks is executed on a
  dedicated core of a Raspberry Pi 3 for ${m \in [1, 10]}$
  threads. For every ${m}$ value, the benchmark is executed 100 times
  totaling 1,000 executions. The maximum \emph{total} cycles
  calculated by Algorithm~\ref{alg:POC-main} from 100 executions of
  ${m}$ threads is recorded as the representative WCET for
  ${m}$ threads. From these WCET values, the minimum
  representative growth factor is calculated for every
  benchmark. 

To verify the benefit of collapse proposed in this work, DAG
  tasks are constructed using the generation pipeline from 
  Section~\ref{sec:evaluation} with one modification: executable
  objects are one of the TacleBench benchmarks with WCET and growth factor
  values estimated by the repeated execution of
  Algortihm~\ref{alg:POC-main}. Nodes within DAG tasks are assigned
  one thread of one benchmark. After assigning 
  executable objects to nodes, the workload ${C_i}$,
  critical path length ${L_i}$, and dedicated cores ${m_i}$ are
  calculated by Equations~\ref{eq:workload},~\ref{eq:criticalpath},
  and~\ref{eq:m}, respectively. The DAG task is then converted to 
  DAG-OT tasks and nodes are collapsed by each of the heuristics. The
  result is four tasks: one DAG requiring ${m_i}$ cores, and three
  DAG-OT requiring ${\hat{m}_i \le m_i}$ cores due to nodes being
  collapsed by the distinct heuristics. The makespan of the four tasks is
  recorded by executing the tasks on the FS platform given the proper
  allocation of ${m_i}$ or ${\hat{m}_i}$ cores. Makespan values are 
  compared to the task's deadline, verifying schedulability and
  illustrating the core savings resulting from collapse.

\subsection{Feasibility Study Results}\label{sec:poc_results}

Growth factors for the 42 benchmarks fall in the range of 0.3 to
7. Benchmarks with a representative growth factor greater than
  1.0 are not collapsed, since they are not beneficial. A sample of
benchmark values are provided in
Table~\ref{table:posgrowthfactorsmall}. The complete list of growth 
factors may be found in the Appendix.

\begin{table}[H]
\centering
\begin{tabular}{r|l|l|l|}
    \cline{2-4}
    Benchmark & fac & matrix1 & ndes \\
    \cline{2-4}
    Growth Factor & 0.42 & 0.84 & 1.38 \\
    \cline{2-4}
\end{tabular}
\caption{Sample TacleBench Benchmarks Growth Factors} 
\label{table:posgrowthfactorsmall}
\end{table}

\begin{table}[h]
  \centering
  \begin{tabular}{|l|r|r|r|r|}
    \hline
    & \multicolumn{4}{c|}{Pre Collapse} \\
    \hline
    $i$ & $C_i$ & $L_i$ & $D_i$ & $m_i$ \\
    \hline
    1 & 6,168,224 & 4,287,924 & 5,248,928 & 3 \\
    2 & 4,616,294 & 3,448,417 & 6,347,882 & 2 \\
    3 & 5,310,666 & 3,614,573 & 3,953,663 & 4 \\
    4 & 6,684,846 & 4,149,946 & 4,448,542 & 4 \\
    \hline
    \hline
    & \multicolumn{4}{c|}{Post Collapse} \\
    \hline
    ${i}$ & $\hat{C}_i$ & $\hat{L}_i$ & $D_i$ & $\hat{m}_i$ \\
    \hline
    1 & 6,087,061 & 5,195,855 & 5,248,928  & 2 \\
    2 & 3,888,725 & 3,888,725 & 6,347,882 & 1 \\
    3 & 5,018,733 & 3,853,186 & 3,953,663 & 3 \\
    4 & 6,342,401 & 3,883,653 & 4,448,542 & 3 \\
    \hline
  \end{tabular}
  \caption{Pre and Post Collapse Metrics}
  \label{table:POCTasks}
\end{table}

A subset of 11 benchmarks were selected as the complete set of
  executable objects when generating tasks. These 11
  were selected based on their growth factors which range from 0.3 to
  2.63. Task graphs are generated with 32 nodes and edge
  probabilities in ${\{0.01, 0.02, 0.03\}}$. Every node is randomly
  assigned one thread of an executable object from the 11
  heuristics. A total of 120 DAG tasks were generated and analyzed
  without collapse and as DAG-OT tasks collapsed by each of the
  benchmarks. Four of 120 were selected  based on their results to
  illustrate the benefits of collapse. The tasks require two, three,
  or four cores. Each task's core allocation is reduced by one to:
  one, two, and three cores respectively. Tasks requiring more cores
  were not considered due to the limited number of Raspberry Pi 3
  devices available.

\begin{figure}[ht]
  \centering
  \includegraphics[width=0.4\textwidth]{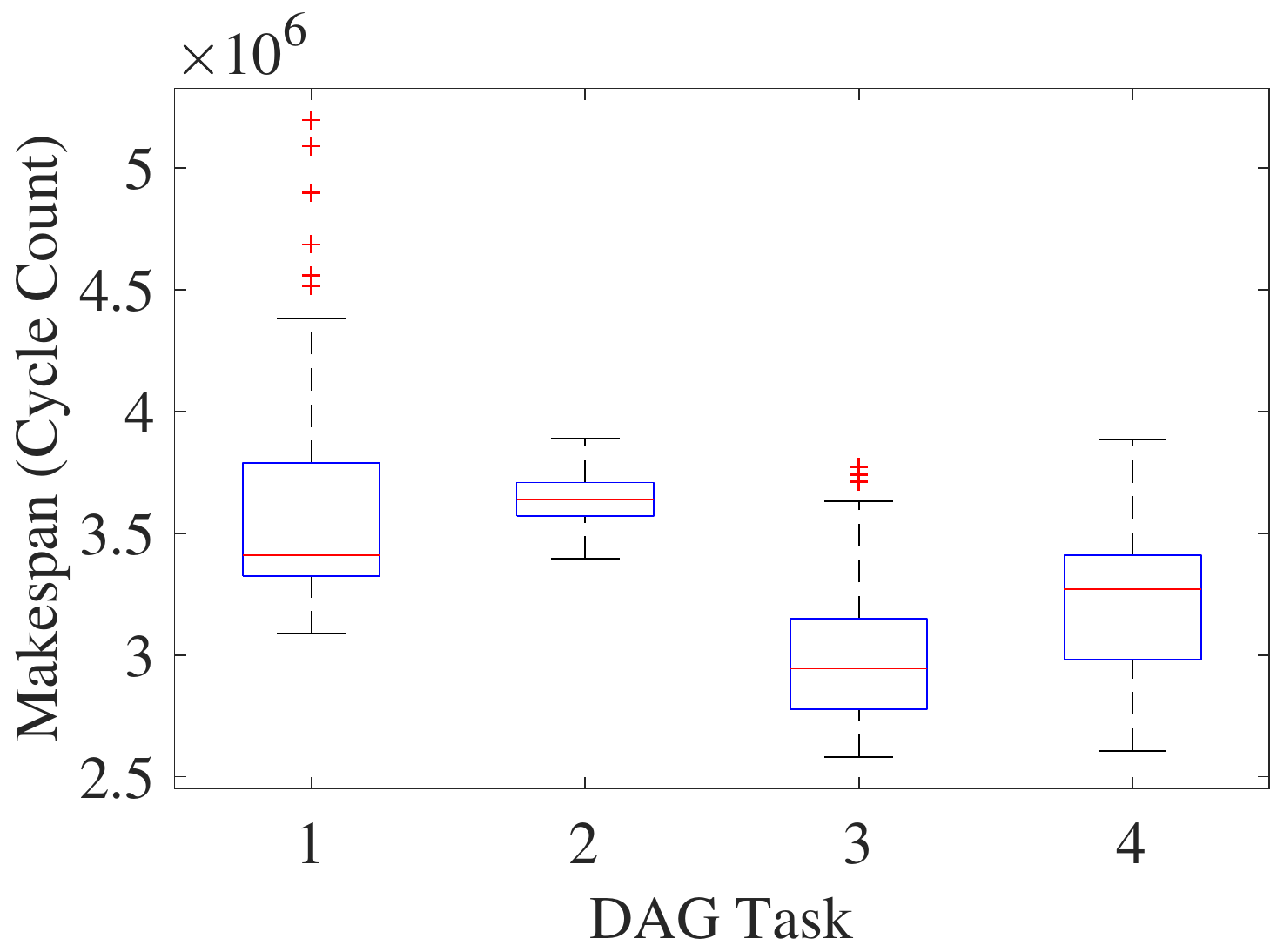}
  \caption{Makespan Distribution for OT-G Collapsed Tasks}
  \label{fig:poc_walltime}
\end{figure}

Table~\ref{table:POCTasks} presents the four selected tasks and the
impact of collapse upon them when run on the POC platform. In this
limited setting, each of the heuristics OT-G, OT-L, and OT-A collapsed the
same set of candidate pairs. Thus, the critical path length and
workload values were similar across all heuristics. The result is a
core savings of 25 to 50 percent.

During execution on the POC, the makespan and workload are recorded.
Given the similar performance of each heuristic,
Figure~\ref{fig:poc_walltime} presents the makespan distribution of
the 100 runs of each task collapsed by OT-G. Variation in makespan
(and workload) may be attributed to interrupts, operating system
interference, or the pseudo-random cache replacement policy. Combined
with Table~\ref{table:pocresults}, the average makespan falls within the
${90^{th}}$ percentile. Given the distribution, the average values are
presented for simplicity.

Table~\ref{table:pocresults} provides the average makespan and
workload savings for all tasks across each of the heuristics. Workload
savings ranges from 2 to 16 percent. The results also verify
schedulability of collapsed task with all makespan values falling
below the deadlines in Table~\ref{table:POCTasks}. In this limited
setting including the negative effect of cache clearing between
threads the savings in makespan, workload, and core allocation are
encouraging for the method proposed herein.
\begin{table}[h]
  \centering
  \begin{tabular}{|l|r|r|r|r|}
    \hline
    & \multicolumn{3}{c|}{Makespan} & \\
    \hline
    $i$ & OT-G     & OT-L    & OT-A     & ${\bar{\Delta}_{C}}$ \\
    \hline 
    1  & 4,531,262  & 5,195,855   & 4,640,880  & 2.03\%		\\
    2  & 3,888,725  & 3,858,028   & 3,942,390  & 16.43\%	\\
    3  & 3,853,186  & 3,600,027   & 3,835,659  & 6.81\%		\\
    4  & 3,883,653  & 4,213,339   & 4,436,712  & 5.12\%		\\
    \hline
  \end{tabular}
  \caption{Mean Makespan and Workload Savings}
  \label{table:pocresults}
\end{table}

\section{Conclusion}
\label{sec:conclusion}

This work proposes the DAG-OT model, joining the federated scheduling
policy and analysis with \bundle{} thread-level scheduling and
analysis through the proposed mechanism of collapsing candidate nodes
of a DAG. The synthetic evaluation and proof of concept support the
proposed mechanism, and heuristic algorithm for selecting and
collapsing nodes; demonstrating the benefit of collapse to
schedulability, workload, and total cores allocated to parallel DAG
tasks. 

There remains an open question of the complexity of optimal
  collapse of a task. Optimal collapse of all tasks within a task set
  remains undefined. The complexity of optimal collapse for a single
  task and all tasks of task set is  reserved for future work. Future
  work also includes consideration for data caches, shared caches
  (evictions and false sharing), and permitting preemptions within
  \bundle{} scheduling.

\clearpage

\appendix{}
To aid in the reproduction of synthetic DAG-OT task sets and support
the proof of concept implementation developed for a network of
Raspberry Pi's, this supplemental material supplies additional details
and results. The first section describes the synthetic taskset
generation pipeline. It differs significantly from other parallel DAG
task set generation mechanisms~\cite{Ueter:2018} due to the inclusion
of infeasible tasks, ie. those tasks where the critical path length is
greater than the relative deadline (${L_i > D_i}$). The second
section provides results from the proof of concept
implementation in greater detail than summarized in the main
work.

\section{Taskset Generation Pipeline}
\begin{figure}[ht]
  \centering
  \includegraphics{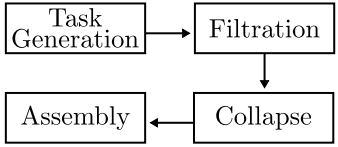}
  \caption{Task Set Generation Pipeline}
  \label{fig:app-eval-pipeline-01}
\end{figure}

Each stage of the pipeline is described using a tuple such as
${\langle A = \{a_1, a_2, ..., a_j\} , B = \{b_1, b_2, ..., b_k\} \rangle}$. A
tuple abbreviates the cross product of all possible combinations
i.e. ${((a_1, b_1), (a_1, b_2), ..., (a_j, a_k))}$. Additionally, a
tuple may be preceded by an iteration constant ${K}$ that repeats each pair
of the cross product ${K}$ times. For example when ${K = 2}$, ${K
\cdot \langle A, B \rangle \rightarrow ((a_1, b_1), (a_1, b_1),
(a_1, b_2), ... , (a_j, b_k), (a_j, b_k))}$. The size of any tuple is
the product of sizes of the elements of the tuple and the iteration
constant.

Task generation is the first stage in the pipeline and is divided into
smaller segments. The first segment of task generation is the creation
of graph structures. There are three input parameters to graph
creation: the number of nodes per graph ${n}$, the probability of an
edge between any two nodes ${P(u,v)}$, and the number iterations
${S}$. To assign an edge to a pair of nodes ${{u,v}}$ a random value
in the range ${r \in [0,1]}$ is generated, if ${r \le P(u,v)}$ the
edge is added. The set of graphs generated is referred to as
${\tau_g}$, which is the result of ${\tau_g = S \cdot \langle n,
  P(u,v) \rangle}$. Table~\ref{table:graph-gen} enumerates these
parameters with a range ${[\min,\max]}$ and increment, the total
provided is the number of graphs generated after this segment.  

\begin{table}[ht]
  \centering
  \begin{tabular}{|c|c|}
    \hline
    Parameter & Range \\
    \hline
    ${n}$ & ${(16,32,64)}$ \\
    ${P(u,v)}$ & ${(0.02, 0.06, 0.12)}$ \\
    ${S}$ & 10 \\
    \hline
    \hline
    Total ${|\tau_g|}$ & 
      ${ |S \cdot \langle n, P(u,v) \rangle| = 90 }$ \\
    \hline
  \end{tabular}

  \caption{Task Generation Graph Creation Parameters}
  \label{table:graph-gen}
\end{table}

The second segment of task generation is execution
assignment. Each task in ${\tau_g}$ is repeatedly assigned objects to
execute, creating a new task after each assignment. Execution
assignment begins by creating a set number of executable objects ${o}$
per task. Each object is given a single thread WCET ${c_1}$, a growth
factor of ${\mathbb{F}}$. The single thread execution value of each
object is assigned a random value from the range ${c_1 \in [1,
    50]}$. The growth factor of each object is assigned a random value
from the range ${[0.2, \mathbb{F}]}$ Every node of the task is
assigned exactly one executable object and one thread of
execution. The set of tasks processed after this segment is referred
to ${\tau_e}$, which is the result of ${\tau_e = \tau_g \times \langle o,
  \mathbb{F} \rangle}$ Table~\ref{table:exe-assign} enumerates the
execution assignment parameters, the total provided is the number of
tasks generated after this segment.

\begin{table}[ht]
  \centering
  \begin{tabular}{|c|c|}
    \hline
    Parameter & Range \\
    \hline
    ${o}$ & ${(4, 8, 16)}$ \\
    ${\mathbb{F}}$ & ${(0.2, 0.6, 1.0)}$ \\
    \hline
    \hline
    Total ${|\tau_e|}$ & 
    {${ |\tau_g \times \langle o, \mathbb{F} \rangle| = 90 \cdot 9 = 810 }$} \\
    \hline
  \end{tabular}
  \caption{Task Generation Execution Assignment Parameters}
  \label{table:exe-assign}
\end{table}

The third and final segment of task generation is timing assignment
for deadlines and periods. Each task in ${\tau_e}$ is repeatedly
assigned a period and deadline, creating a new task after each
assignment. Timing assignment is related to the \emph{task target
utilization} values ${U_\tau}$. For each task target utilization
value, the task's period is set to ${T = C / U_\tau}$. Since all tasks
have implicit deadlines ${D = T}$. The set of tasks after
task set generation is referred to as ${\tau}$, which is the result of
${\tau = \tau_e \times \langle U_\tau \rangle}$. After which, the
set of tasks ${\tau}$ is sent to
filtration. Table~\ref{table:timing-assign} enumerates the timing
assignment parameters and provides the total number of tasks
generated. 

\begin{table}[ht]
  \centering
  \begin{tabular}{|c|c|}
    \hline
    Parameter & Range \\
    \hline
    ${U_\tau}$ & ${(0.25, 0.50, 2.0, 4.0, 8.0, 16.0)}$ \\
    \hline
    \hline
    Total ${|\tau|}$ &
    {${ |\tau_e \times \langle U_\tau \rangle| = 810 \cdot 6 =
  4,860 }$} \\
    \hline
  \end{tabular}
  \caption{Task Generation Timing Assignment Parameters}
  \label{table:timing-assign}
\end{table}

Filtration is a single step process that removes tasks that are
\emph{always} trivially infeasible. A trivially infeasible task has a
critical path length greater than its deadline ${(L_i >
D_i)}$ \textbf{or} the number 
of allocated cores exceeds the number of nodes in the task (${m_i
> V_i}$). Since collapse may reduce the critical path length of a DAG
task, an infeasible task may become a feasible DAG-OT task. Filtration
executes each of the collapse heuristics on every task of ${\tau}$. If
the DAG task of ${\tau_i}$ is feasible, the task remains. If the DAG
task is infeasible, and any collapse ordering produces a feasible
DAG-OT task, the task remains. Only if the DAG task is infeasible, and
all collapse orderings are also infeasible is the task removed from
${\tau}$. It should be noted that any post-collapse DAG-OT task
${\hat{\tau}_i}$ which is trivially infeasible could not have
originated from a pre-collapse trivially infeasible DAG task
${\tau_i}$. Additionally, there may be one or more trivially
infeasible DAG tasks ${\tau_i}$ that could be collapsed into a
feasible DAG-OT task ${\hat{\tau}_i}$ if the optimal collapse order
was known. Such optimally collapsed feasible tasks are omitted from
the evaluation, for they would contribute negatively equally to the
existing DAG and DAG-OT schedulability tests.

Collapse is the next stage of the pipeline, for each DAG task in
${\tau}$ a collapsed version of the DAG-OT task is produced. Tasks are
segregated into pools one for the DAG task, and one for each of the
collapse orders applied to the DAG-OT task. These collapsed task sets
are referred to as ${\tau_a}$ for arbitrary ordering, ${\tau_b}$
for greatest benefit, and ${\tau_p}$ for least penalty. Each DAG task
${\tau_i \in \tau}$ shares its index ${i}$ across pools, for example:
${\tau_i \in \tau_p}$ refers to the DAG-OT task generated from the DAG
task ${\tau_i \in \tau}$ that was collapsed using the least penalty
heuristic. 

Assembly is the final stage of the task set generation
pipeline. Fore every selection of cores in the system architecture
${c}$, and \emph{target task set utilization} ${U}$, ${N}$ task sets
are created from the DAG tasks ${\tau}$. For every task set assembled
from ${\tau}$, the corresponding task set from each of the collapse
orderings is also created. To clarify, for a DAG task set ${A =
  \{\tau_i, \tau_j, \tau_k\}, \tau_i,\tau_j,\tau_k \in \tau}$, the
corresponding task DAG-OT task set using the greatest benefit collapse
ordering is ${A_b = \{\tau_i, \tau_j, \tau_k\}, \tau_i,\tau_j,\tau_k
\in \tau_b}$. Table~\ref{table:assembly} enumerates the assembly
parameters and the total number of task sets created. The total
reflects the total number of DAG task sets assembled, it does not
reflect the equivalent DAG-OT task sets.

\begin{table}[ht]
  \centering
  \begin{tabular}{|c|c|}
    \hline
    Parameter & Range \\
    \hline
    ${U}$ & ${(0.5, 1, 2, 4, 8, 12, 16, 20, 24, 28, 32, 36)}$ \\
    ${c}$ & ${(4, 8, 12, 16, 20, 24, 28, 32)}$ \\
    ${N}$ & 1,000 \\
    \hline
    \hline
    Total & 
    {${ N \cdot \langle c, U \rangle = 96,000}$} \\
    \hline
  \end{tabular}
  \caption{Task Set Assembly Parameters}
  \label{table:assembly}
\end{table}

\subsection{TacleBench Growth Factors}

For each of the TacleBench benchmarks the rate at which their observed
WCET grows with respect to the number of threads executed serially was
collected. This value is referred to as the growth factor of the
benchmark presented in Table~\ref{table:GrowthFactor}. The observed
WCET is the greatest number of cycles observed from 100 runs per
thread value. The thread values range from one to ten, for a total of
1,000 runs per benchmark.  

\begin{table}[th]
\centering
\begin{tabular}{|c|c|c|}
\hline
Program         & Growth Factor \\
\hline
Ammunition      & 1.001017384   \\
Binary Search   & 0.438874249   \\
Bitcount        & 0.591137743   \\
Bitonic         & 2.487587942   \\
Bsort           & 0.973991342    \\
cjpeg\_transupp & 1.003582809   \\
cjpeg\_wrbmp    & 0.800890651   \\
Complex Updates & 0.712378749   \\
Cosf            & 2.181066416   \\
Count Negative  & 2.633100341   \\
Cubic           & 1.001100335   \\
Deg2Rad         & 3.510920367   \\
Dijkstras       & 0.999209534   \\
Epic            & 0.975179889   \\
Fac             & 0.415288078    \\
FFT             & 0.939265074   \\
FilterBank      & 0.990190452   \\
Fir2Dim         & 3.412174094   \\
Fmref           & 0.904876397   \\
IIR             & 7.053002576   \\
InsertSort      & 7.422310306   \\
Isqrt           & 1.013520967   \\
Jfdctint        & 0.682709413   \\
Lift            & 0.97349589   \\
Lms             & 1.036344984   \\
Ludcmp          & 0.300562783   \\
Matrix1         & 0.844821934  \\
MD5             & 1.001594396   \\
Minver          & 0.662344871   \\
MPEG2           & 0.998725266   \\
Ndes            & 1.376412296   \\
Petrinet        & 3.859254005   \\
PM              & 0.996212215   \\
PowerWindow     & 0.941866543   \\
QuickSort       & 0.972317214   \\
Rad2Deg         & 3.099475684   \\
Recursion       & 0.744205596   \\
Rinjdael\_dec   & 0.989430449   \\
Rinjdael\_enc   & 0.982374859   \\
Sha             & 0.995654855    \\
St              & 1.06888128   \\
Statemate       & 0.906628026   \\
Susan           & 0.996663067   \\
\hline
\end{tabular}
\caption{Growth Factors for TacleBench Benchmarks}
\label{table:GrowthFactor} 
\end{table}

\clearpage
\bibliography{cache,dag}

\end{document}